\documentclass[10pt]{IEEEtran}
\usepackage{float}
\usepackage{graphicx}
\usepackage{clrscode}
\usepackage{stfloats}
\usepackage{paralist}
\usepackage{tikz}
\usepackage{soul} 
\usepackage{color}
\usepackage{cite}
\usepackage{multicol}
\usepackage{tabularx}
\usepackage{comment}
\usepackage{amsmath, amssymb, mathrsfs, amsfonts}
\usepackage{amsthm}
\usepackage{mathtools}
\usepackage{epsfig, epstopdf}
\usepackage{algorithm}
\usepackage{enumitem}
\usepackage{algorithmic}
\usepackage{lipsum}
\usepackage{pbox}
\usepackage{array}
\usepackage{tablefootnote}
\usepackage{subfigure}

\usetikzlibrary{arrows,automata}

\allowdisplaybreaks

\newtheorem{lemma}{Lemma}

\newtheorem{remark}{Remark}

\usepackage{hyperref}

\newcommand{\RNum}[1]{\uppercase\expandafter{\romannumeral #1\relax}}
\begin{document}

\title{Massive MIMO-NOMA Systems Secrecy in the Presence of Active Eavesdroppers}

\author{
		\IEEEauthorblockN{Marziyeh Soltani, Mahtab Mirmohseni, \textit{Senior Member, IEEE}, and Panos Papadimitratos, \textit{Fellow, IEEE} \\
		\vspace*{0.5em}
			}\thanks{Part of this paper has been presented at the International Conference on Computer Communications and Networks
(ICCCN) 2021 \cite{inproceedings}.\\
M. Soltani and M.Mirmohseni are with 5/6GIC, the Institute for Communication Systems (ICS),
University of Surrey, GU2 7XH Guildford, U.K. (e-mail: m.soltani@surrey.ac.uk, m.mirmohseni@surrey.ac.uk). P. Papadimitratos is with Networked Systems Security group, KTH Royal Institute of Technology, Sweden (e-mail: papadim@kth.se).}}

\maketitle
\begin{abstract}
Non-orthogonal multiple access (NOMA) and massive multiple-input multiple-output (MIMO) systems are highly efficient. Massive MIMO systems are inherently resistant to passive attackers (eavesdroppers), thanks to transmissions directed to the users. However, active attackers can transmit a combination of legitimate user pilot signals during the channel estimation phase. This way, they can mislead the base station (BS) to rotate the transmission in their direction and allow them to eavesdrop during the downlink data transmission phase. In this paper, we analyze this vulnerability with two-user pairing strategies and investigate how physical layer security can mitigate such attacks and ensure secure (confidential) communication. We derive the secrecy outage probability (SOP) and a lower bound on the ergodic secrecy capacity using stochastic geometry tools when the number of antennas in the BSs tends to infinity (i.e., massive MIMO BS). The numerical and simulation results show that one of the strategies performs better and has a higher ergodic secrecy rate (ESR) and lower SOP. Moreover, we show that using NOMA instead of orthogonal multiple access (OMA) improves system performance significantly.
\end{abstract}
\section{Introduction}\label{introduction}
Massive multiple-input multiple-output (MIMO) and non-orthogonal multiple access (NOMA) are two promising technologies envisioned for beyond 5G and 6G \cite{Larsson2014MassiveMF, Ding2014ontheperfo}, increasing the spectral and energy efficiency, with the help of base stations (BSs) with large antenna arrays supporting many users in the same frequency-time domain \cite{ZhenYang2017EnergyEfficient}. 
NOMA separates users in the power domain and allows operating simultaneously in the same frequency, by exploiting superposition coding at the transmitter and successive interference cancellation (SIC) at the receivers \cite{HIgu2015NOMASIC}. Additionally, it is employed in various technologies such as unmanned aerial vehicles (UAVs) \cite{uplinkprecoding,energyefficient}. Combining the advantages of massive MIMO and NOMA is being investigated \cite{zhang2017fullynoma,zhang2018exploit}. For a large number of users in each cell, clustering users with the same pilot sequence is beneficial. With a limited number of orthogonal pilot sequences, the residual interference after imperfect SIC is reduced.
However, user clustering (and use of the same pilot) degrades the uplink training, decreasing spatial resolution, because of intra-cluster pilot interference.

As wireless networks increasingly transmit sensitive information, ensuring their security is crucial. While massive MIMO offers protection against passive eavesdropping through large antenna arrays and focused beamforming, it remains vulnerable to active eavesdroppers, especially when they transmit the same pilot sequences as legitimate users during the channel estimation phase. Thus, examining the combination of massive MIMO and NOMA in the presence of active eavesdroppers, while accounting for the inherent randomness in wireless networks, is vital for securing future communication systems, including those anticipated for 6G networks.

Existing research on the impact of active eavesdroppers can be categorized into three main areas: first, studies focusing on detecting the presence of active eavesdroppers in the system \cite{DetectionofactiveeavesdroppersinmassiveMIMO, DetectionofPilotContaminationAttackbasedonUncoordinatedFrequencyShifts, PilotContaminationAttackDetection}; second, works analyzing system performance under the influence of attackers, which includes this work; and third, studies addressing methods to mitigate the impact of active eavesdroppers \cite{mitigate1, mitigate2}. Focusing on three efforts in analyzing security for massive MIMO NOMA systems, there are significant gaps. Existing works on massive MIMO NOMA either overlook security constraints \cite{Kusalad2018ratemassivnoma, zhang2017fullynoma, out2} or focus on a single-cell scenario without considering the random locations of users \cite{zhang2018exploit}.

Our contribution addresses this latter gap by analyzing
the impact of active eavesdroppers in a multicell massive
MIMO NOMA network, where BSs, attackers, and users are
distributed according to independent homogeneous Poisson
point processes (HPPPs). The attackers interfere with the channel estimation phase by sending a combination of pilot sequences. We note that the standard pilot contamination in massive MIMO systems primarily degrades the channel estimation phase. However, in the adversarial scenario, the focus of this work, there is an additional effect beyond the channel estimation error and the resultant degradation in the legitimate user’s rate: information leakage to the active attackers during the data transmission phase, while in the absence of a pilot-phase attack, the information leakage to the eavesdropper is negligible due to the highly directional nature of beamforming in massive MIMO systems. We
quantify the information leakage to adversaries and derive
performance limits using both ergodic secrecy capacity and
SOP as metrics for fast and slow fading scenarios, respectively.
 Through simulations, we identify the conditions under which physical layer security (PLS) techniques can counteract such attackers. To achieve this, we have overcome several key challenges:

$\bullet$ Channel Estimation Complexity:  During the channel estimation phase, the received pilot signals from users at the BS are affected by multiple channels: the channels of users within the same cluster and cell, the channels of users in different clusters within the same cell, the channels of users in the same cluster across different cells, the channels of users in different clusters across different cells, and the attackers channels. Thus, the estimated channels at the BS become complicated and depend on multiple channels. The BS then uses this complicated channel estimation to design the precoder for the downlink data, resulting in increased complexity in deriving the signal-to-interference-plus-noise Ratio (SINR) due to correlated channel vectors.

$\bullet$ Randomness in Wireless Networks: The inherent randomness in large-scale fading (path loss) and small-scale fading (Rayleigh fading) complicates the analysis of key performance metrics, such as secrecy outage probability (SOP) and ergodic secrecy rate (ESR), making it difficult to derive precise results.

$\bullet$ Pairing Strategies in NOMA: NOMA introduces correlations between the distances of users within a cluster, which complicates the derivation of the distribution of these distances. Additionally, it makes the analysis of the statistical properties of inter-cell interference and information leakage to eavesdroppers more challenging.

Our contribution addresses the  aforementioned challenges and achieves the following:

$\bullet$ We provide a connection-level analysis and derive asymptotic SINR expressions for legitimate users and attackers as the number of BS antennas approaches infinity in a massive MIMO scenario. We consider imperfect channel estimation under a worst-case adversarial scenario (i.e., the strongest attacker).

$\bullet$ We analyze the system secrecy performance by deriving exact expressions for ESR and SOP for arbitrary user pairs. Each user connects to the nearest BS, and BSs select a fixed number of users in their Poisson-Voronoi (PV) cell to serve, clustering them into two NOMA users. We examine two pairing strategies: random-pairing (RP) \cite{Kusalad2018ratemassivnoma}, where users are paired based on normalized distance, and interference-based-pairing (IBP) \cite{new1}, where users are paired based on proximity to both the serving BS and the closest BS among all other BSs. We show that RP guarantees SIC for the central user, while in IBP, SIC is sometimes performed by the second user, though with a negligible probability. We emphasize that the focus of this paper is not to introduce new pairing strategies but to use existing approaches to pair NOMA users, which allows us to further analyze the system performance metrics under each scheme in a random massive MIMO NOMA system. Moreover, we derive closed-form expressions for the lower bounds of the ergodic rates of both the central and second users in the RP and IBP schemes.

$\bullet$ Methodological Innovations: We apply the law of large numbers to manage small-scale fading randomness and use order statistics for large-scale fading randomness, enabling us to derive the distance distribution. We utilize two methods for deriving ESR and SOP for both central and secondary users in RP and IBP: one introduces a dummy gamma random variable for applying the Alzer inequality  \cite[Appendix A]{Bai2015alzer}, and the other is based on Laplace transforms for computational efficiency. We also utilize the probability generating functional (PGFL) and the Slivnyak-Mecke theorem \cite{book} to analyze intra-cell interference and correlated terms, simplifying the expressions with a Taylor series expansion.

$\bullet$ In the Simulation results, we compare the performance of IBP and RP, both internally and in comparison to OMA. The simulation and numerical results demonstrate that the two methods—one based on Alzer's inequality and the other on the Laplace transform—used to derive the ESR and SOP for both pairing strategies (RP and IBP) closely match in all figures, thereby validating the accuracy of each method. We analyze how factors such as the power coefficient of active eavesdroppers, BS density, eavesdropper density, NOMA power coefficient, and target rate affect SOPs and ESRs.

In the rest of the paper, Section~\ref{related} discusses related works, and Section~\ref{model} describes the system and adversary model. Section~\ref{strategy} details the transmission strategy. In Sections~\ref{ergodic_rate} and \ref{SOP}, we characterize the ESR and SOP of the system, respectively. Our simulation results are provided in Section~\ref{simulations} before we conclude.
\section{Related Works}\label{related}
Regarding the security of NOMA, in \cite{PhysicalLayerSecurityforNOMASystemsRequirements}, the authors provide a comprehensive overview of current PLS-aided NOMA systems. Various scenarios are explored, such as those involving active and passive eavesdroppers, and combinations with relay and reconfigurable intelligent surfaces (RISs). In \cite{OnSecureNOMASystemsWithTransmitAntenna}, the authors examine the secrecy performance of a downlink NOMA system with a multi-antenna BS, two legitimate receivers, and an eavesdropper, considering both single-input and single-output (SISO) and multiple-input and single-output (MISO) systems with different transmit antenna selection strategies. \cite{SecrecyOutageofMax–MinTASScheme} investigates secure communications in a single-cell MIMO NOMA system, assuming Nakagami-m fading and multiple antennas at all nodes. In \cite{polisetti2021physical}, the BS communicates with users using NOMA while an eavesdropper intercepts confidential information. \cite{hu2019physical} aims to improve secrecy probability using friendly jammers in a NOMA network, employing a stochastic geometry approach to analyze SOP. \cite{Liu2017enhancing} studies a network with users uniformly distributed around the BS and eavesdroppers according to an HPPP, deriving exact SOP expressions for single- and multiple-antenna scenarios. \cite{Abol2019onsecnoma} focuses on a downlink NOMA network with two users and an external passive eavesdropper, deriving SOP and ESR in closed form by considering channel coefficient randomness.

Regarding massive MIMO security, \cite{Akgun2019vulnerability} examines a single-cell massive MIMO system where active eavesdroppers aim to minimize the downlink sum rate, considering precise and probabilistic locations of BS and users. In \cite{Wu2016securemassive}, an asymptotic achievable secrecy rate is derived for a multi-cell massive MIMO system with fixed-location users and a multi-antenna attacker as the number of BS antennas approaches infinity. \cite{zhang2018exploit} explores a single-cell massive MIMO-NOMA network with an active eavesdropper in each cluster, deriving a closed-form ESR expression considering channel coefficient randomness.

Massive MIMO and NOMA combinations without security constraints have been studied in various scenarios. \cite{zhang2017fullynoma} designs a fully non-orthogonal communication system for massive access, deriving a tight lower bound on spectral efficiency considering channel coefficient randomness. \cite{Kusalad2018ratemassivnoma} examines a massive MIMO-NOMA network, obtaining the achievable rate of a typical user with imperfect SIC for randomly located users and BSs. \cite{out2} derives the outage probability (OP) and bit error rate for a single-cell downlink massive MIMO-NOMA system using random matrix theory.

\textbf{Comparison to existing work:} The most relevant works are \cite{Kusalad2018ratemassivnoma, zhang2018exploit, OnSecureNOMASystemsWithTransmitAntenna, SecrecyOutageofMax–MinTASScheme}. [9] and [15] consider massive MIMO-NOMA scenarios. Unlike [9], which derives the ergodic secrecy rate (ESR) for a single-cell massive MIMO-NOMA system with an active eavesdropper in each cluster—considering only small-scale fading randomness—we derive both the secrecy outage probability (SOP) and ESR by accounting for the randomness of both small scale and \textit{large-scale fading} (as a function of user locations) in a \textit{multi-cell} system. In a multi-cell scenario, the relative locations of all nodes significantly impact the performance metrics, adding complexity to the derivations. Specifically, multi-cell interference must be considered in the rates of both users and eavesdroppers, in addition to intra-cluster interference caused by users within the same cluster. Moreover, instead of assuming exactly one eavesdropper per cluster, we \textit{randomly distribute} eavesdroppers throughout the network—an assumption that better reflects realistic deployments. Moreover, in contrast to [15], which studies a massive MIMO-NOMA network without addressing security, our work offers a \textit{security-focused investigation.} Furthermore, unlike [9] and [15], we examine \textit{two distinct user-pairing} strategies (RP and IBP) to derive the performance metrics under each strategy. Also, while [15] characterizes intercell interference using moment matching (approximating the interference distribution by matching mean and variance to an inverse Gaussian distribution), we derive the \textit{exact distributions} of all interference terms.
\\
Furthermore, [21] and [22] consider secrecy performance of a downlink conventional NOMA systems \textit{rather than massive MIMO-NOMA}. Specifically, [21] and [22] focus on secrecy in a single-cell NOMA scenario, primarily considering small-scale fading (Rayleigh and Nakagami-$m$) and assuming perfect channel state information (CSI) at the BS. Additionally, these studies deal with passive eavesdroppers. In contrast, our work considers a multi-cell setting, incorporating both intra-cell and inter-cell interference, as well as spatially correlated terms. We also introduce large-scale fading and derive our results using stochastic geometry, requiring modeling of the point process that governs distances between interfering BSs, users, and eavesdroppers. Our analysis explicitly includes a channel estimation phase and active eavesdroppers that can interfere during this phase. Channel estimation is thus critical in evaluating the impact of active eavesdropping.
The first part of the results is presented in \cite{inproceedings}, which includes the SOP and ESR analysis for RP. In addition, in this paper, we now derive SOP and ESR for IBP. we also provide simpler expressions for results of \cite{inproceedings}. Moreover, for both RP and IBP, we utilize another method based on the Laplace transformation to derive SOP and ESR, in addition to the previous method based on the Alzer inequality. Additionally, the simulation and numerical section is completely novel.

\begin{figure}
\centering
\includegraphics[width=3.5in]{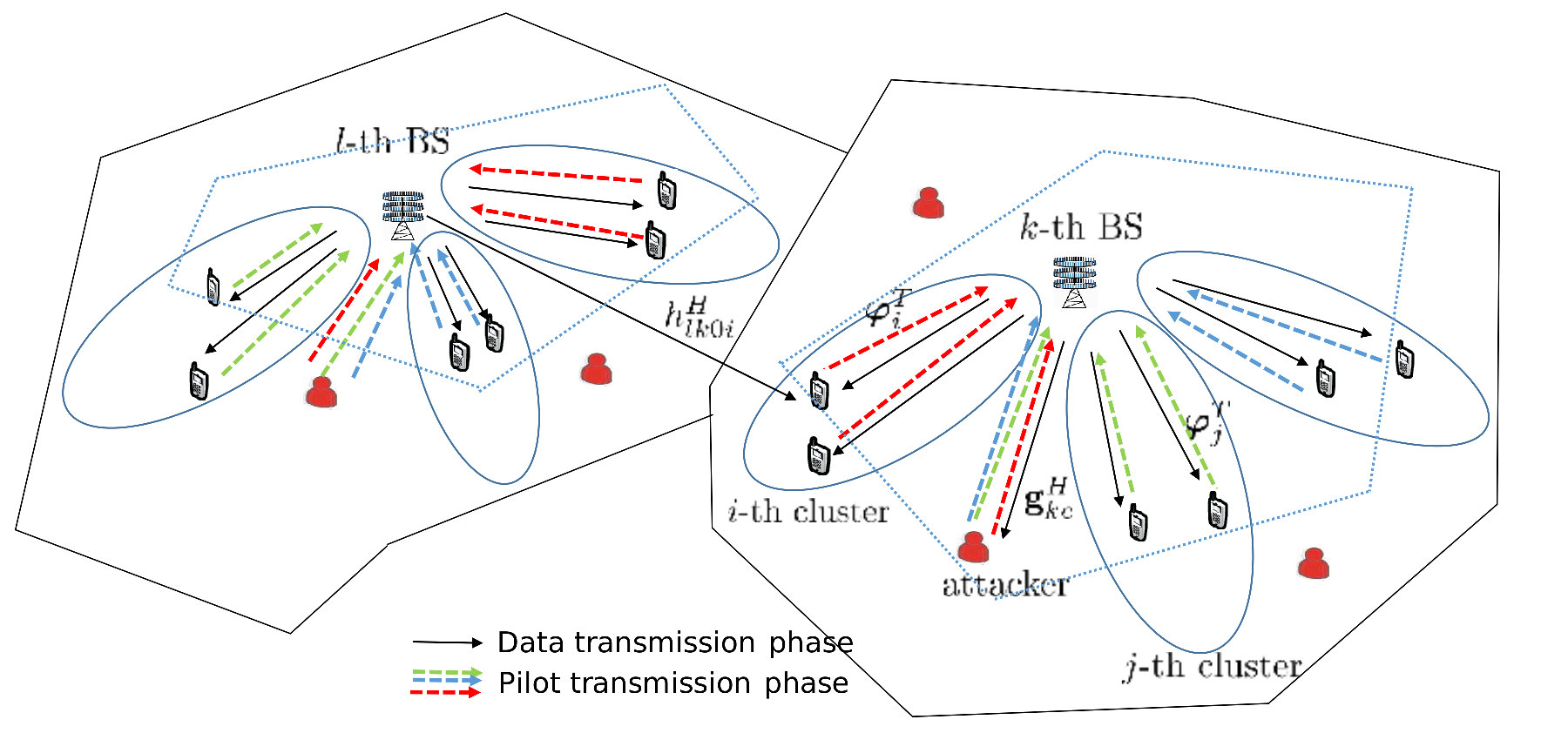}
\caption{System model: Dashed and solid lines show the pilot and data transmission phases, respectively. For the RP strategy, users are randomly selected from the solid area, and in each cluster, the closer user (based on normalized distance) is chosen as the central user. For the IBP strategy, the central users are randomly selected from the dotted area and the second users are randomly selected from the area outside the dotted area and inside the solid area. The dotted area is different for each user and it is formed based on the distance between the BS serving the user and the closest BS to the serving BS.}\label{model1}
\end{figure}

\textbf{Notation:} Bold letters denote vectors. $P(.), f_x(.)$, $F_x(.),$ $ E[.]$ represent the probability, the probability density function (PDF), the CDF, and the expectation, respectively. $\mathbf{X}^T, \mathbf{X}^H$ and $\mathbf{X}^*$ are the transpose, the Hermitian transpose, and the conjugate of $X$, respectively. $\mathbf{I}_M$ is an $M\times M$ identity matrix. The Euclidean norm is $\parallel. \parallel$ and $\mathcal{CN}(.,.)$; denotes a multi-variate circularly-symmetric complex Gaussian distribution; $\mathbb{C}$ and $\mathbb{N}$ are sets of complex and natural numbers. $\Gamma(\cdot)$ refers to the Gamma function. Important notations and symbols used in this work are given in Table \ref{tab:symbols}.
\begin{table}[ht]
\caption{\textcolor{blue}{List of symbols and their definitions}}
\centering
\renewcommand{\arraystretch}{1.2}
\begin{tabularx}{\columnwidth}{|p{1.8cm}|X|}
\hline
\textbf{Symbol} & \textbf{Definition} \\
\hline
\( \Phi_b \) & \textcolor{blue}{HPPP for BSs with density \( \lambda_b \).} \\
\hline
\( \Phi_e \) & \textcolor{blue}{HPPP for eavesdroppers with density \( \lambda_e \).} \\
\hline
\( M \) & \textcolor{blue}{Number of antennas at each BS.} \\
\hline
\( I \) & \textcolor{blue}{Number of clusters in each cell.} \\
\hline
\( \alpha \) & \textcolor{blue}{Path loss exponent.} \\
\hline
\( \mathbf{h}_{klmi} \) & \textcolor{blue}{Small-scale fading channel vector between the user and the BS.} \\
\hline
\( r_{klmi} \) & \textcolor{blue}{Distance between the user and the BS.} \\
\hline
\( \mathbf{g}_{ke} \) & \textcolor{blue}{BS-eavesdropper small-scale fading channel vector.} \\
\hline
\( r_{ke} \) & \textcolor{blue}{Distance between the \( k \)-th BS and an eavesdropper \( e \).} \\
\hline
\( \boldsymbol{\varphi}_i \) & \textcolor{blue}{Pilot sequence assigned to cluster \( i \).} \\
\hline
\( P_p \) & \textcolor{blue}{Maximum transmission power during the pilot phase.} \\
\hline
\( P_d \) & \textcolor{blue}{Downlink transmission power.} \\
\hline
\( a_m \) & \textcolor{blue}{Power coefficient allocated to user \( m \) for downlink transmission.} \\
\hline
\( b_m \) & \textcolor{blue}{Power coefficient for user \( m \) during the pilot phase.} \\
\hline
\( d_i \) & \textcolor{blue}{Power allocation coefficient for attackers, \( i \)-th pilot sequence.} \\
\hline
\( \sigma^2 \) & \textcolor{blue}{Noise variance.} \\
\hline
\( R \) & \textcolor{blue}{Sum of power coefficients: \( R = \sum_{m=0}^{1} \frac{a_m}{b_m} \).} \\
\hline
\( S_m \) & \textcolor{blue}{Normalized distance metric: \( S_m = \frac{r_{kkmi}^{-2\alpha}}{\sum_{l \neq k} r_{lkmi}^{-2\alpha}} \).} \\
\hline
\( \tau \) & \textcolor{blue}{IBP boundary threshold.} \\
\hline
\( \tilde{R}_0 \) & \textcolor{blue}{Target data rate for central user.} \\
\hline
\( \tilde{R}_1 \) & \textcolor{blue}{Target data rate for second user.} \\
\hline
\( N \) & \textcolor{blue}{Shape parameter of Gamma random variable.} \\
\hline
\( U \) & \textcolor{blue}{Shape parameter for Alzer inequality.} \\
\hline
\( \eta \) & \textcolor{blue}{Derived from \( N \) and Gamma function.} \\
\hline
\( \tilde{\eta} \) & \textcolor{blue}{Derived from \( U \) and Gamma function.} \\
\hline
\end{tabularx}
\label{tab:symbols}
\end{table}
\section{System Model}\label{model}
The system model is illustrated in Fig. \ref{model1}.
We consider a multi-cell NOMA-enabled massive MIMO system operating in TDD mode for downlink transmission. BSs are distributed according to an HPPP denoted as $\mathrm{\Phi}_\text{b}\cup\lbrace0\rbrace$, with a density of $\lambda_b$. Each cell contains a BS equipped with $M$ antennas and multiple single-antenna users distributed uniformly and independently over their serving BS’s PV cell, which is the region closer to that BS than to other BSs, in a fully loaded scenario. Due to Slivnyak's theorem \cite{book}, the original measure of $\mathrm{\Phi}_\text{b}$ is equal to the reduced palm measure of $\mathrm{\Phi}_\text{b}\cup\lbrace0\rbrace$ \cite{book}. Thus, without loss of generality, we assume the $k$-th BS is located at the origin, and we analyze users chosen from this BS's PV cell. Additionally, single-antenna non-colluding active eavesdroppers are distributed according to another HPPP, denoted as $\mathrm{\Phi}_\text{e}$, with a density of $\lambda_e$.

In TDD mode, two phases occur during a coherence time interval: pilot transmission (channel estimation) and downlink data transmission. During the pilot phase, users transmit pilot sequences to allow the BS to estimate the channel coefficient vectors. Based on these estimates, the BS calculates precoding vectors for each user, which are then utilized during the downlink transmission of data. In the pilot phase, each user establishes a connection with its closest BS. Subsequently, each BS selects a predetermined number of $2I$ users, within its respective PV cell, to provide service to.

We employ a two-user NOMA, although our results can be extended to $m$-user NOMA. The $2I$ users in a cell are organized into $I$ clusters. We assume two different approaches for clustering, the RP and IBP strategies, which will be explained later, to select and organize the users into 2-user NOMA clusters \footnote{For the sake of simplicity, we assume an even number of users to be \textit{selected} for clustering into two-user NOMA, regardless of the number of the users available in the cell. This is a common assumption in NOMA scenarios \cite{nomaclustering1, nomaclustering2}. However, if an odd number of users is selected to be served, one user who is not assigned to any cluster can be served using OMA instead of NOMA, as suggested in \cite{nomaoma}.}. For both the RP strategy and the IBP strategy, each cluster designates one user as central and the other as the second user based on some distances. Note that all internal users are considered legitimate.

Distinct, orthogonal pilot sequences are assigned to each cluster in a cell, while users have the same pilot sequence within a cluster. If $\boldsymbol{\varphi}_i$ and $\boldsymbol{\varphi}_j$ are the pilot sequences assigned to cluster $i$ and $j$, respectively, then we have:
\begin{eqnarray}
\boldsymbol{\varphi}^H_i \boldsymbol{\varphi}_j=\left \{ \begin{array} {rl} 1 \qquad & i=j \\
0 \qquad &\mathrm{otherwise}
\end{array}, \right. \label{pilot}
\end{eqnarray}
where $\boldsymbol{\varphi}$ is a $q\times 1$ vector. The same cluster-pilot mapping is used for all cells. As the maximum number of mutually orthogonal sequences of length $q$ is equal to $q$, the number of clusters in each cell, $I$, must not be greater than $q$, to guarantee orthogonality \footnote{In a vector space of dimension \( q \), the maximum number of linearly independent and mutually orthogonal sequences (or basis vectors) is at most \( q \). If \( I > q \), then at least one of the pilot sequences \( \boldsymbol{\varphi} \) must either be reused or be a linear combination of other mutually independent pilot sequences. This contradicts the assumption of mutual independence in the pilot codebook (eq. (1)).}. We assume that all active eavesdroppers know the pilot codebook (same in all cells) and send a combination of all pilot sequences during the channel estimation phase. As a result, during the downlink transmission, they can eavesdrop data intended for any user in any cluster. We investigate the worst case by considering an active eavesdropper that has the maximum SINR.

In our channel model, we account for both small-scale fading and large-scale path loss effects. Our channel model excludes shadowing effects and comprises two distinct categories:
\\
i) The channel vector between the $m$-th user in the $i$-th cluster of the $l$-th cell and the BS in the $k$-th cell is denoted by:
\begin{align}
\mathbf{h}_{klmi}r_{klmi}^{-\alpha/2},\quad \mathrm{~}\mathbf{h}_{klmi}\sim\mathcal{CN}(\mathbf{0},\mathbf{I}_M),
\end{align}
where $m \in \lbrace 0, 1\rbrace$, $i \in \lbrace 1,...,I\rbrace$, $k$, $l \in \mathbb{N}$, $\mathbf{h}_{klmi} \in \mathbb{C}^{M\times 1}$ represents the small-scale fading coefficient, $r_{klmi}$ is the distance between the user and the $k$-th BS, and $\alpha$ is the path loss exponent. We assume uncorrelated Rayleigh fading with no dominant spatial directivity \cite{correlation}. According to the law of large numbers:
\begin{eqnarray}
\frac{\mathbf{h}^H_{klmi}\mathbf{h}_{\hat{k} \hat{l} \hat{m} \hat{i}}}{M}\overset{M\rightarrow\infty}=
\left \{ \begin{array} {rl}1 &k=\hat{k}, l=\hat{l}, m=\hat{m}, i=\hat{i} \\
0 &\mathrm{otherwise}
\end{array}.\right.	\label{law1}
\end{eqnarray}
\\
ii) The channel vector between the $k$-th BS and an arbitrary active eavesdropper is denoted by:
\begin{align}\mathbf{g}_{ke}r_{ke}^{-\alpha/2},\quad
\mathrm{~}\mathbf{g}_{ke}\sim\mathcal{CN}(\mathbf{0},\mathbf{I}_M),
\end{align}
where $\mathbf{g}_{ke} \in \mathbb{C}^{M\times 1}$ is the small-scale fading coefficient, and $r_{ke}$ is the distance between the active eavesdropper and the $k$-th BS. According to the law of large numbers, $\mathbf{h}^H_{klmi}\mathbf{g}_{\hat{k} e}/M$ tends to 1 in probability as $M\rightarrow\infty$, only when the active eavesdropper and the legitimate user are at the same location. Otherwise, it tends to be zero. Similarly, for two illegitimate channel vectors, we have:
\begin{eqnarray}
\frac{\mathbf{g}^H_{ke}\mathbf{g}_{\hat{k} \tilde{e}}}{M}\overset{M\rightarrow\infty}=
\left \{ \begin{array} {rl}1 &k=\hat{k} ,e=\tilde{e} \\
0 &\mathrm{otherwise}
\end{array}. \right. \label{law2}
\end{eqnarray}
\\
We next present the two pairing strategies.

\textbf{Random-pairing (RP) strategy:}
Each BS selects $2I$ users, two users for each of the $I$ clusters, from its connected users located within its respective PV cell based on the following approach: The BS randomly selects two users from its connected users to form the first cluster, designating them as cluster $1$. The first pilot sequence from the pilot codebook is assigned to this cluster. For the second cluster, the BS again randomly selects two users from the remaining connected users, distinct from those in cluster $1$. The second pilot sequence from the codebook is allocated to cluster $2$. The BS repeats this procedure until $I$ clusters are established. The full load scenario assumption for the users ensures that, within the pairing strategy, the selection of any user to be placed in a cluster does not affect the point process of other users, and that the PDF of the user-BS distance remains unchanged. Within each cluster, NOMA users are categorized as either ``central'' or ``second'' based on their normalized distance from the serving BS. Specifically, users in the $i$-th cluster of the $k$-th cell, $i \in \lbrace 1,...,I\rbrace$ and $k \in \mathbb{N}$, are ordered based on descending $S_m$, defined as:
\begin{align}
S_m=\frac{r_{kkmi}^{-2\alpha}}{\sum_{l\neq k}^{}r_{lkmi}^{-2\alpha}},\label{sm}
\end{align}
where $m\in\lbrace 0,1\rbrace$.
In each cluster, we term a user \textit{central} if it has $\max(S_0, S_1)$. The other user is termed the \textit{second} user. In Section V, we show that the central users in RP are always able to perform SIC.

\textbf{Interference-based-pairing (IBP) strategy:} 
Each BS selects $2I$ users, two users in each of the $I$ clusters, from its connected users located within its respective PV cell based on the following approach: within the PV cell of each BS, we divide the region into two distinct areas, and each user in these regions is classified based on its distances from both the serving BS and the dominant interfering BSs. To clarify, if a user is located at position $y$ within the PV cell of a BS positioned at $X$, denoted as $\Upsilon_{X}=\lbrace y\in \mathbb{R}^2| \parallel y-X \parallel\leq  \parallel y-x \parallel | x \in \phi_b \rbrace$, we identify it as central if it belongs to the set $\Upsilon_{Xc}=\lbrace y \in \Upsilon_{X}| \parallel y-X \parallel\leq \min_{x\in \mathrm{\Phi}\text{b}\setminus X} \parallel y-x \parallel\tau\rbrace$, where $\tau\in(0,1)$ represents the boundary threshold. Conversely, if the user falls within the set $\Upsilon_X\setminus \Upsilon_{Xc}$, we categorize it as the second user. After forming the regions, the BS randomly selects one user from the region $\Upsilon_{Xc}$ and another user from the region $\Upsilon_X\setminus \Upsilon_{Xc}$, designating them as cluster $1$. These two users are assigned the first pilot sequence from the pilot codebook. For the second cluster, the BS once again randomly chooses a user from the region $\Upsilon_{Xc}$ different from the cluster one's user and selects another user at random from the region $\Upsilon_X\setminus \Upsilon_{Xc}$ distinct from the cluster one's user. The BS repeats this process until forming $I$ clusters. Similar to RP, the full load scenario assumption for the users ensures us that the selection of any user to be placed in a cluster does not affect the point process of other users. Later, we show that there are some conditions under which the central users cannot perform SIC; however, the second user can. This is because we categorize the users as central and second based on the function of distances, not on the ability to perform SIC. Moreover, we demonstrate that the probability of this event is nearly zero, and almost always the central users can perform SIC.

We aim to derive the fundamental limits of performance metrics in RP and IBP by considering both ESR and SOP as suitable metrics in fast and slow fading scenarios, respectively.
\section{Transmission phase for the RP and IBP strategy}\label{strategy}
\subsection{Channel Estimation Phase}
Each BS estimates downlink channel vectors to design precoding vectors for the data transmission phase. Users in cluster $i$ send pilot $\boldsymbol{\varphi}^T_i$, with $b_i$ the user power allocation coefficient. $P_p$ is the maximum transmission power for each user and attacker. Each attacker sends $\sum_{i=1}^{I}\sqrt{d_iP_p}\boldsymbol{\varphi}^T_i$, in which $d_i$ is the power allocation coefficient for the $i$-th pilot sequence. The received signal at the end of the pilot transmission phase at the $k$-th BS is:
$\mathbf{y}_k=\sum_{m=0}^{1}\sum_{i=1}^{I}\sum_{l=1}^{\infty}\mathbf{h}_{klmi}r_{klmi}^{-\alpha/2}
\boldsymbol{\varphi}^{T}_i\sqrt{b_mP_p}+\sum_{e\in\ \mathrm{\Phi}_\text{e}}\mathbf{g}_{ke}r^{-\alpha/2}_{ke}(\sum_{i=1}^{I}\sqrt{d_iP_p}\boldsymbol{\varphi}^T_i)+\mathbf{W}_k,$ where $m$, $i$ and $l$ denote, respectively, the $m$-th user in the $i$-th cluster of the $l$-th cell; $\mathbf{W}^{M\times q}_k$ is the noise component at the $k$-th BS with i.i.d elements drawn according to a distribution $ \mathcal{CN}(0,\sigma^2) $. In $\mathbf{y}_k$, the first term denotes the pilot sequences sent by all legitimate users and the second term denotes the pilot sequences sent by the attackers. To estimate the channel vector of a legitimate user, the BS multiplies the received signal, $\mathbf{y}_k$, with the user pilot sequence. This eliminates the effect of the other pilot sequences that are by design orthogonal. However, the effect of the pilot sequences sent by the users in the same cluster as the mentioned user remains. The effect of the adversarial transmission remains too, as it was a combination of all pilot sequences.

By defining $\mathcal{B}=\lbrace (\hat{m},l)\vert \hat{m}\in\lbrace 0,1\rbrace , l \in \mathbb{N}\rbrace$, the estimation of the channel vector for the $m$-th user in the $\tilde{i}$-th cluster in the $k$-th cell, expressed at the $k$-th BS, is:
\begin{align}
\hat{\mathbf{h}}_{kkm\tilde{i}}\overset{(a)}{=}\frac{\mathbf{y}_k\boldsymbol{\varphi^*}_{\tilde{i}}}{\sqrt{b_mP_p}}
&\overset{(b)}=\mathbf{h}_{kkm\tilde{i}}r^{-\alpha/2}_{kkm\tilde{i}}
+\!\!\!\!\!\!\!\!\!\!\!\sum_{\mathcal{B}\setminus \lbrace \hat{m}=m, l=k\rbrace}\!\!\!\!\!\!\!\!\frac{\mathbf{h}_{kl\hat{m} \tilde{i}}r_{kl\hat{m} \tilde{i}}^{-\alpha/2}\sqrt{b_{\hat{m}}}}{\sqrt{b_m}}
\nonumber\\&+\sum_{e\in\mathrm{\Phi}_\text{e}}\frac{\mathbf{g}_{ke}r^{-\alpha/2}_{ke}\sqrt{d_{\tilde{i}}}}{\sqrt{b_m}}+\frac{\mathbf{W}_k\boldsymbol{\varphi^*}_{\tilde{i}}}{\sqrt{b_mP_p}},\label{estimat}
\end{align}
where (a) follows from the above estimation method and (b) follows from (\ref{pilot}) and $\mathbf{y}_k$.
\subsection{Downlink Data Transmission Phase}\label{orderingdef}
For the precoding design, in massive MIMO systems, linear precoding is nearly optimal \cite{Rusek2013scaling}; we use here one such scheme, matched filter precoding. We use different precoding vectors for each user in a cluster, based on their estimated channels. The $l$-th BS multiplies the intended data for user $m$ in the $i$-th cluster of the $l$-th cell by $ \hat{\mathbf{h}}_{llmi} $. In the rest of the paper, $m=0$ denotes the central user and $m=1$ denotes the second user. The power coefficient allocated to user $m$ for downlink transmission is $a_m$; thus, $a_0+a_1=1 $. BS allocates power $P_da_m$ to user $m$. The transmitted power is normalized by $2M$ (the number of users in each cluster multiplied by the number of BS antennas). $s_{lmi}$ is the data the $l$-th BS sends to user $m$ in the $i$-th cluster of the $l$-th cell, with $\parallel s_{lmi} \parallel=1$. $w_d$ is scalar additive white Gaussian noise with variance $\sigma^2_n$. $\mathbf{x}^{M\times 1}_l$ the signal transmitted by the $l$-th BS is \footnote{$l$ represents the number of cells in our system.}:
\begin{align}
\mathbf{x}_l=\sum_{i=1}^{I}\sum_{m=0}^{1}\hat{\mathbf{h}}_{llmi}s_{lmi}\sqrt{\frac{P_da_m}{2M}}.\label{x_l}
\end{align}\\
\textit{1) Received signal at the legitimate user:} Now, we derive the received signal at an arbitrary legitimate user. $\mathbf{h}_l$ is the channel coefficient and $r_l$ is distance between the user and the $l$-th BS. The received signal at the legitimate user is \footnote{When using stochastic geometry to analyze multicell systems, a common assumption is that the number of BSs (cells) is infinite, as per \cite{elahe, Kusalad2018ratemassivnoma, Network-LevelIntegratedSensingandCommunication, StochasticGeometryBasedModelingandAnalysisofMassive, Stochasticgeometricmodelingandinterferenceanalysis, ImpactofArtificialNoiseonCellularNetworksAStochasticGeometryApproach}.}:
\begin{align}
y=\sum_{l=1}^{\infty}\mathbf{h}_l^H r_l^{-\alpha/2}\mathbf{x}_l+w_d.\label{sig}
\end{align}
By substituting (\ref{estimat}) and (\ref{x_l}) into (\ref{sig}) and by defining $\tilde{\mathcal{B}}=\lbrace (s,t)\vert s\in\lbrace 0,1\rbrace , t\in \mathbb{N}\rbrace$, the received signal at the central user in the $\tilde{i}$-th cluster of the $k$-th cell is expressed by replacing $y$, $\mathbf{h}_l$ and $r_l$ with $y_{k0\tilde{i}}$, $\mathbf{h}_{lk0\tilde{i}}$ and $r_{lk0\tilde{i}}$ in (\ref{sig}). Similarly, the received signal at the second user in the $\tilde{i}$-th cluster of the $k$-th cell, $y_{k1\tilde{i}}$, can be expressed by substituting (\ref{estimat}) and (\ref{x_l}) into (\ref{sig}) and replacing $y$, $\mathbf{h}_l$ and $r_l$ with $y_{k1\tilde{i}}$, $\mathbf{h}_{lk1\tilde{i}}$ and $r_{lk1\tilde{i}}$, respectively.\\
\textit{2) Received signal at the attacker:} The signal received at an arbitrary attacker, $\tilde{e}$, can be expressed by substituting (\ref{estimat}) and (\ref{x_l}) into (\ref{sig}), and by replacing $y$, $\mathbf{h}_l$ and $r_l$ with $y_{\tilde{e}}$, $\mathbf{g}^H_{l\tilde{e}}$ and $r_{l\tilde{e}}$. The attacker experiences the same noise power as the legitimate users. Now, we let $M \rightarrow \infty$, and we use (\ref{law1}) and (\ref{law2}). Based on the approach in Appendix~\ref{lemma1p}, noting $ \sum_{i=1}^{I}d_i=1$ and ignoring $\sigma^2
_n$, the SINR of the attacker eavesdropping, $w_0$, the message to the central user, is:
\begin{align}
\mathrm{SINR}^{w_0}_{\tilde{e}}=\frac{\frac{d_{\tilde{i}}a_0r^{-2\alpha}_{k\tilde{e}}}{b_
0}}{r^{-2\alpha}_{k\tilde{e}}(R-\frac{d_{\tilde{i}}a_0}{b_0})+R\sum_{l\neq k}^{\infty}r^{-2\alpha}_{l\tilde{e}}}.\label{jam}
\end{align}
We remark that, unlike the central user, the attacker cannot perform SIC, thus being subject to comparatively increased interference. Moreover, from (\ref{jam}), it is clear that if the attacker had not sent its pilot sequence during the channel estimation phase (i.e., $d_{\tilde{i}}=0$), it would not have received any information, due to the massive MIMO directed beam. If the attacker intends to eavesdrop the second user message ($w_1$), its SINR is derived similarly by substituting $a_0$ and $b_0$ with $a_1$ and $b_1$, respectively.
\section{Ergodic Secrecy Rate}\label{ergodic_rate}
We derive a lower bound on the ESR for our system model (Section \ref{model}). The ESR of a legitimate user can be computed as $C_s=E[(R_{\textrm{user}}-R_{\textrm{eavs}})^+]$, where $(x)^+=\max(0,x)$ and $R_{\textrm{user}}$ and $R_{\textrm{eavs}}$ are the rates of the legitimate user and the attacker downlink channel, respectively.
As the BS is equipped with a massive number of antennas, the ESR is lower-bounded by\cite{zhu2014securema} $C_s=(E[R_{\textrm{user}}]-E[R_{\textrm{eavs}}])^+$.
We compute the achievable rates for the legitimate and the eavesdropper channels separately.
\subsection{RP strategy}
Now, to derive the achievable ESRs, we have to obtain $ \mathrm{SINR}^{w_0}_{k0\tilde{i}}$, the SINR of the central user (in the $\tilde{i}$-th cluster of $k$-th cell) when decoding its own message ($w_0$) after omitting the second user's message ($w_1$); $ \mathrm{SINR}^{w_1}_{k0\tilde{i}}$, the SINR of the central user when decoding the second user signal; and $ \mathrm{SINR}^{w_1}_{k1\tilde{i}}$, the SINR of the second user (in $\tilde{i}$-th cluster of $k$-th cell) when decoding its own signal. In Lemma \ref{lemma1} (proof at Appendix \ref{lemma1p}), we derive these SINRs, and by using them in Lemma \ref{lemma2} (proof at Appendix \ref{lemma2p}), we derive the ergodic rates (ERs) of the users. We use two methods for deriving ERs in Lemma \ref{lemma2}. Method 1 is based on the Alzer inequality (all equations containing the shape parameter \(N\)), while method 2 relies on the Laplace transformation. In Section \ref{simulations}, we will demonstrate that both methods are close to each other and Monte Carlo simulations. Finally, we express the ergodic leakage rate to the strongest attacker in Lemma \ref{lemma3} (proof at Appendix \ref{lemma3p}).
\begin{lemma}\label{lemma1}
$\mathrm{SINR}^{w_0}_{k0\tilde{i}}$, $\mathrm{SINR}^{w_1}_{k0\tilde{i}}$ and $\mathrm{SINR}^{w_1}_{k1\tilde{i}}$ are obtained as following:
\begin{align}
\mathrm{SINR}^{w_0}_{k0\tilde{i}}=&\frac{a_0r^{-2\alpha}_{kk0\tilde{i}}}{Rb_0\sum_{l\neq k}^{\infty}r^{-2\alpha}_{lk0\tilde{i}}},\label{hello}\\
 \mathrm{SINR}^{w_1}_{k1\tilde{i}}=&\frac{a_1r^{-2\alpha}_{kk1\tilde{i}}}{\frac{a_0b_1r^{-2\alpha}_{kk1\tilde{i}}}{b_0}+Rb_1\sum_{l\neq k}^{\infty}r^{-2\alpha}_{lk1\tilde{i}}}.\label{hello2}\\
\mathrm{SINR}^{w_1}_{k0\tilde{i}}=&\frac{\frac{b_0a_1r^{-2\alpha}_{kk0\tilde{i}}}{b_1}}{Rb_0\sum_{1\neq k}^{\infty}r^{-2\alpha}_{lk0\tilde{i}}+a_0r^{-2\alpha}_{kk0\tilde{i}}},\label{hello22}
\end{align}
\end{lemma}
\begin{lemma}\label{lemma2}
The ER of the central and the second user in the $\tilde{i}$-th cluster of the $k$-th cell, denoted by $R_0$ and $R_1$, are derived as:

Method 1:
\begin{align}
&R_0=\!\!\!\!\int_{0}^{\infty}(1-(1-\sum_{n=1}^{N}\frac{(-1)^{n+1}{N\choose{n}}(-\alpha+1)}{-\alpha+1-R\eta n(\frac{(2^t-1)b_0}{a_0})
})^2)dt\label{R_Df}\\
&R_1=\!\!\!\!\int_{0}^{\log(1+\frac{a_1b_0}{b_1a_0})}\!\!\!\!(\sum_{n=1}^{N}\frac{(-1)^{n+1}{N\choose{n}}(-\alpha+1)}{-\alpha+1-R\eta n(\frac{(2^t-1)b_1}{a_1-\frac{(2^t-1)a_0b_1}{b_0}})
})^2dt\label{R_Df2}
\end{align}
Method 2:
\begin{align}
&R_0=\!\!\int_{0}^{\infty}\!\!1-e^{2(1-\alpha)(\frac{a_0}{(2^t-1)Rb_0})}dt.\\
&R_1=\!\!\int_{0}^{\log{\frac{a_1b_0}{a_0b_1}+1}}
\!\!\!\!\!\!\!(1-e^{(1-\alpha)(\frac{a_1-\frac{(2^t-1)a_0b_1}{b_0}}{(2^t-1)Rb_1})})^2dt.
\end{align}
Moreover, lower bounds on these ERs are derived as:
\begin{align}
&R_0 \geq \log_2\left(1 + \frac{2(\alpha-1)a_0}{R b_0}\right).\\
&R_1 \geq \log_2\left(1 + \frac{a_1}{\frac{a_0 b_1}{b_0} + \frac{3 R b_1}{2(\alpha - 1)}}\right)
\end{align}
\end{lemma}
\begin{lemma}\label{lemma3}
The ergodic leakage rate to the attacker that seeks to eavesdrop data of the central user in the $\tilde{i}$-th cluster of $k$-th cell is derived as follows:
\begin{align}
R^0_e&=\!\!\!\!\int_{0}^{\infty}\!\!\!\frac{1-\exp(-2 \pi \lambda_e[\sum_{u=1}^{U}(-1)^{u+1}{U\choose{u}}\frac{e^{-xu\tilde{\eta} (\frac{Rb_0}{d_{\tilde{i}}a_0}-1)}}{2(\frac{\pi \lambda_b x u \tilde{\eta} R b_0}{(\alpha-1)d_{\tilde{i}}a_0})}])}{\ln{2}(1+x)}dx\\
&\approx \frac{1}{\ln 2} \int_0^{\frac{\bar{\eta}}{1 - \bar{\eta}}} \frac{1 - \exp\left(-\frac{\lambda_e}{\lambda_b} \left(\frac{\bar{\eta}- x (1 - \bar{\eta})}{x}\right)^{1/\alpha} \frac{1}{\Gamma(1 + 1/\alpha)}\right)}{1 + x} \, dx.
\label{e55}
\end{align}
where \( \bar{\eta} = \frac{d_{\tilde{i}} a_0}{R b_0} \). Similarly, the ergodic leakage rate to the attacker seeking to eavesdrop the second user data is obtained by replacing $a_0$ and $b_0$ with $a_1$ and $b_1$ in (\ref{e55}).
\end{lemma}
The ESR of the central and second user can be computed by using Lemma \ref{lemma2} and Lemma \ref{lemma3}.
\begin{remark}
Obtaining closed-form expressions for Lemma \ref{lemma2} and Lemma \ref{lemma3} has proven to be challenging. However, the derived expressions are exact and can be numerically evaluated using a single integral function in MATLAB. Additionally, as demonstrated in Appendix \ref{lemma2p}, the inner terms of the integrals correspond to the complements of certain CDFs for which we have derived closed forms. Thus, the integral limit can be reduced to the value at which the CDF reaches 1, which depends on the system parameters. For example, as illustrated in Fig. \ref{cdf0} (a), for $\alpha=4$, $F_{S_0}(s_0)$ ($F_{\mathrm{SINR}^{w_0}_{\tilde{e}}}(x)$) reaches almost 1 at approximately 500 (0.5). Consequently, the integral limit can be reduced to $t<\log_2(\frac{500a_0}{Rb_0}+1)$ at $R_0$ (or 0.5 at $R^0_e$).
\end{remark}

In the following, we provide an intuitive validation of the theoretical findings in Lemmas 2 and 3, demonstrating that these derivations are consistent with expected results. If we set \(a_0 = 0\) in $R_0$ at Lemma \ref{lemma2}, the ER of the central user becomes zero, as expected, since no power is directed toward this user. Similarly, setting \(a_1 = 0\) in $R_1$ at Lemma \ref{lemma2}, results in the ER of the second user being zero. Furthermore, by setting \(a_0 = 0\) in (\ref{e55}) and utilizing \( \lim_{a_0 \rightarrow 0} \frac{e^{-xu\tilde{\eta} (\frac{Rb_0}{d_{\tilde{i}}a_0}-1)}}{2(\frac{\pi \lambda_b x u \tilde{\eta} R b_0}{(\alpha-1)d_{\tilde{i}}a_0})} = 0 \), the ER of eavesdropping on the central user also becomes zero. Additionally, the ERs of the users are independent of the BS density. This independence arises from our assumption of a fully loaded scenario for the users, coupled with directed beamforming in massive MIMO systems. By setting \(a_0 = a_1\) and \(b_0 = b_1\), the ERs of the users become independent of these power coefficient parameters, and we have \(R_0 > R_1\). This is because our NOMA user pairing is based on distance rather than power coefficients. Moreover, the ERs of the users are independent of the cell and cluster numbers, \( k \) and \( \tilde{i} \). However, the ER of the eavesdropper depends on the specific cluster it is targeting. For example, if the power dedicated to that cluster, denoted as \( d_{\tilde{i}} \), is zero, the ER of the eavesdropper will also be zero.
\subsection{IBP strategy}\label{newlabeladd}
For the IBP strategy, $S_0$ and $S_1$, defined in Section \ref{model}, do not directly affect the selection of the central and second users. However, as discussed in APPENDIX \ref{lemma2p}, if \(S_0>S_1\), then \(\mathrm{SINR}^{w_1}_{k0\tilde{i}} > \mathrm{SINR}^{w_1}_{k1\tilde{i}}\) and the central user can perform SIC; otherwise, it cannot. As a result, we need to calculate the ER conditioned on the two events, $S_0>S_1$ and $S_0<S_1$. First, in Lemma \ref{lemma1new} (proof at Appendix \ref{lemma1newp}), we derive the CDF of these two random variables. Then, for $S_0>S_1$ ($S_0<S_1$),
we use two methods for deriving ERs of the central and the second users in Lemma \ref{lemma2new} (Lemma \ref{lemma3new}) (proof in Appendix \ref{lemma2newp} (Appendix \ref{lemma3newp})): method 1, which is based on the Alzer inequality and the result of Lemma \ref{lemma1new}, and method 2, which is based on the Laplace transformation.
\begin{lemma}\label{lemma1new}
By defining $W_i(r_m,r_{kki\tilde{i}})=r_mr_{kki\tilde{i}}e^{-\pi \rho \lambda_b r_m^2-\eta ns_ir_{kki\tilde{i}}^{2\alpha}r_m^{-2\alpha}+\frac{\pi \lambda_b\eta n s_i r_{kki\tilde{i}}^{2\alpha}r_m^{-2\alpha+2}}{1-\alpha}}$ for $i=0,1$, the CDF of $S_0$ and $S_1$ are obtained as following;
\begin{align}
&F_{S_0}\!(s_0)\!\!=1\!\!+\!\!\!\!\sum_{n=1}^{N}\!\!
\frac{(-1)^{n}\!{N\choose{n}}(2\pi \rho \lambda_b)^2}{\tau^2}\!\!\!\!\int_{0}^{\infty}\!\!\!\!\!\int_{0}^{\tau r_m}\!\!\!\!\!\!\!\!\!\!\!\!W_0(r_m,r_{kk0\tilde{i}})dr_m\!\!dr_{kk0\tilde{i}}\nonumber\\
&= 1 + \sum_{n=1}^{N} \frac{(-1)^n \binom{N}{n} (2\pi \rho \lambda_b)^2}{\tau^2} \cdot \frac{1}{4\alpha} \int_{0}^{\tau^{2\alpha}} \frac{w^{\frac{1-\alpha}{\alpha}} e^{-B w}}{(A - B_0 C w)^2} dw.\label{cdfs0new}\\
&F_{S_1}\!(s_1)\!\!=1\!\!+\!\!\!\!\sum_{n=1}^{N}\!\!
\frac{(-1)^{n}\!{N\choose{n}}(2\pi \rho \lambda_b)^2}{1-\tau^2}\!\!\!\!\int_{0}^{\infty}\!\!\!\!\!\int_{\tau r_m}^{r_m}\!\!\!\!\!\!\!\!\!W_1(r_m,r_{kk1\tilde{i}})dr_m\!dr_{kk1\tilde{i}}\nonumber\\
&=1 + \sum_{n=1}^{N} \frac{(-1)^n \binom{N}{n} (2\pi \rho \lambda_b)^2}{1-\tau^2} \frac{1}{4\alpha} \int_{\tau^{2\alpha}}^{1} \frac{w^{\frac{1-\alpha}{\alpha}} e^{-B w}}{(A - B_1 C w)^2} dw,
\label{cdfs1new}
\end{align}
where $A = \pi \rho \lambda_b$, $B_i = \eta n s_i$, $C = \frac{\pi \lambda_b}{1-\alpha}$. The CDFs $F_{S_1}(s_1)$ and $F_{S_0}(s_0)$ under the IBP strategy are shown in Fig. \ref{cdf0}(a). The second equalities in (\ref{cdfs0new}) and (\ref{cdfs1new}), representing the equivalent simplified forms (“equ. sim. form”), are also illustrated in Fig. \ref{cdf0}(a).
\begin{figure}
    \centering
    \subfigure[$F_{S_0}(s_0)$, $F_{S_1}(s_1)$, $F_{\tilde{S_0}}(\tilde{s_0})$, and $F_{\tilde{S_1}}(\tilde{s_1})$, at RP and IBP,  $F_{\mathrm{SINR}^{w_1}_{\tilde{e}}}(s)$ and $F_{\mathrm{SINR}^{w_1}_{\tilde{e}}}(s)$ at eaves.]{\includegraphics[scale=.29]{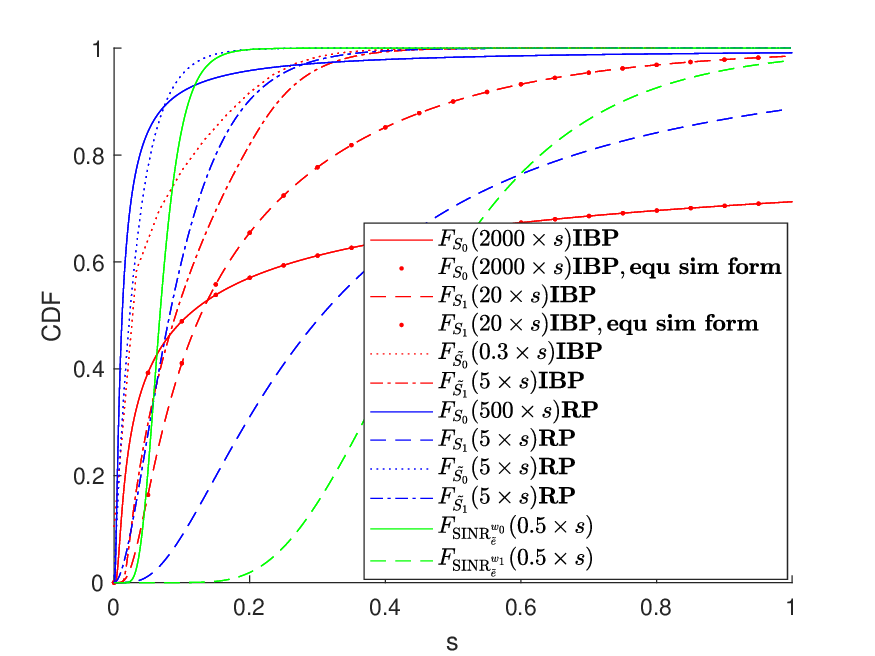}}\label{a}
    \subfigure[$f_{\mathrm{SINR}^{w_0}_{\tilde{e}}}(x)$ and $f_{\mathrm{SINR}^{w_1}_{\tilde{e}}}(x)$, $a_0=0.1$, $b_0=0.4$]{\includegraphics[scale=.29]{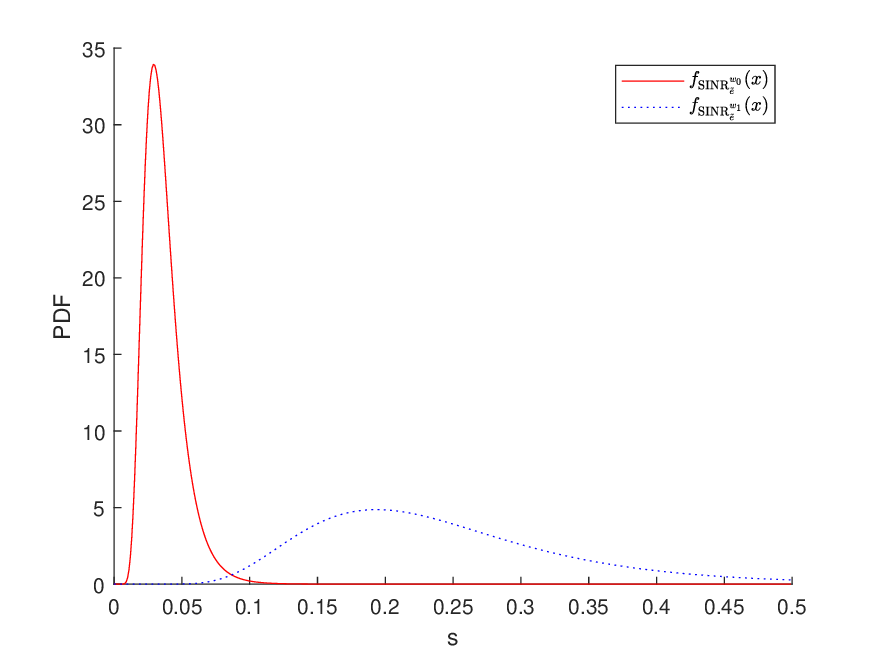}}\label{b}
    \caption{CDF and PDF, $\lambda_b=5\times 10^{-6.5}$, $\lambda_e=5\times 10^{-6.5}$, $N=7$, $\tau=0.7$, and $\alpha=4$.}
    \label{cdf0}
\end{figure}
\end{lemma}
\begin{lemma}\label{lemma2new}
If $S_0>S_1$, the ER of the central and the second users in the $\tilde{i}$-th cluster of $k$-th cell are denoted by $R_{a0}$ and $R_{a1}$ respectively and are derived as follows:

Method 1:
\begin{align}
&R_{a0}=\!\!\int_{0}^{\infty}\!\!\!(1-\!\!F_{S_0}(\frac{(2^t-1))Rb_0}{a_0}))dt.\\
&R_{a1}\!\!=\!\!\int_{0}^{\log{\left(\frac{a_1b_0}{a_0b_1}+1\right)}}\!\!\!\!\!\!\!\!\!\!\!(1-F_{S_1}(\frac{(2^t-1)Rb_1}{a_1-\frac{(2^t-1)a_0b_1}{b_0}}))dt.
\end{align}
 where $F_{S_0}(s_0)$ and $F_{S_1}(s_1)$ are given at Lemma \ref{lemma1new}.
 
Method 2:
\begin{align}
\!\!&R_{a0}=\!\!\frac{(2\pi \rho \lambda_b)^2}{\tau^2}\!\!\!\!\int_{0}^{\infty}\!\!\!\!\!\int_{0}^{\tau r_m}\!\!\!\!\!\!\!\!\!\!\!r_mr_{kk0\tilde{i}}e^{-\pi \rho \lambda_b r_m^2}\!\tilde{N}(r_{kk0\tilde{i}},r_m)dr_mdr_{kk0\tilde{i}}\nonumber\\
 &R_{a1}\!\!=\frac{(2\pi \rho \lambda_b)^2}{1-\tau^2}\!\!\!\!\int_{0}^{\infty}\!\!\!\!\!\int_{\tau r_m}^{r_m}\!\!\!\!\!\!\!\!\!r_mr_{kk1\tilde{i}}e^{-\pi \rho \lambda_b r_m^2}\tilde{M}(r_{kk1\tilde{i}},r_m)dr_mdr_{kk1\tilde{i}}.
\end{align}
where $\tilde{N}(r_{kk0\tilde{i}},r_m)\triangleq\log_{2}(\frac{a_0}{r_{kk0\tilde{i}}^{2\alpha}r_m^{-2\alpha}(1-\frac{\pi \lambda_b r_m^{2}}{1-\alpha})Rb_0}+1)$ and $\tilde{M}(r_{kk1\tilde{i}},r_m)\triangleq \\
  \min(\log_{2}(\frac{a_1}{r_{kk1\tilde{i}}^{2\alpha}r_m^{-2\alpha}(1-\frac{\pi \lambda_b r_m^{2}}{1-\alpha})Rb_1+\frac{b_1a_0}{b_0}}+1),\log_{2}({\frac{a_1b_0}{a_0b_1}+1}))$.
Moreover, lower bounds on these ERs are derived as $R_{a0} \geq \log_2(1 + \frac{a_0 (1+\alpha) }{Rb_0\tau^{2\alpha} })$ and $R_{a1} \geq \log_2(1 + \frac{a_1}{\frac{a_0 b_1}{b_0} + R b_1 \left[\tau^{2\alpha} + \frac{\alpha}{\alpha + 1} \left(1 - \tau^{2(\alpha + 1)/\alpha}\right)\right]})$.
\end{lemma}
\begin{lemma}\label{lemma3new}
If $ S_0<S_1$, the ER of the central and the second users in the $\tilde{i}$-th cluster of $k$-th cell are denoted by $R_{b0}$ and $R_{b1}$ respectively and are derived as follows:

Method 1:
\begin{align}
&R_{b0}=\!\!\!\!\int_{0}^{\log{\frac{a_0b_1}{a_1b_0}+1}}\!\!\!\!\!\!\!(1-F_{S_0}(\!\frac{(2^t-1)Rb_0}{a_0-\frac{(2^t-1)a_1b_0}{b_1}}))dt\nonumber\\
&R_{b1}=\int_{0}^{\infty}(1-F_{S_1}(\frac{(2^t-1)Rb_1}{a_1}))dt
\end{align}
where $F_{S_0}(s_0)$ and $F_{S_1}(s_1)$ are given at Lemma \ref{lemma1new}.

Method 2:
\begin{align}
&R_{b0}\!\!=\!\frac{(2\pi \rho \lambda_b)^2}{\tau^2}\!\!\!\!\int_{0}^{\infty}\!\!\!\!\!\int_{0}^{\tau r_m}\!\!\!\!\!\!\!\!\!\!\!\!r_mr_{kk0\tilde{i}}e^{-\pi \rho \lambda_b r_m^2}\tilde{M_2}(r_{kk0\tilde{i}},\!r_m)dr_mdr_{kk0\tilde{i}}\nonumber\\
&R_{b1}\!\!=\!\frac{(2\pi \rho \lambda_b)^2}{1-\tau^2}\!\!\!\!\int_{0}^{\infty}\!\!\!\!\!\int_{\tau r_m}^{r_m}\!\!\!\!\!\!\!\!\!r_mr_{kk1\tilde{i}}e^{-\pi \rho \lambda_b r_m^2}\tilde{N_2}(r_{kk1\tilde{i}},\!r_m)dr_mdr_{kk1\tilde{i}.}
\end{align}
where $\tilde{N_2}(r_{kk1\tilde{i}},r_m)\triangleq\log_{2}(\frac{a_1}{r_{kk1\tilde{i}}^{2\alpha}r_m^{-2\alpha}(1-\frac{\pi \lambda_b r_m^{2}}{1-\alpha})Rb_1}+1)$ and $\tilde{M_2}(r_{kk0\tilde{i}},r_m)\triangleq \\ \min(\log_{2}(\frac{a_0}{r_{kk0\tilde{i}}^{2\alpha}r_m^{-2\alpha}(1-\frac{\pi \lambda_b r_m^{2}}{1-\alpha})Rb_0+\frac{b_0a_1}{b_1}}+1),\log_{2}({\frac{a_0b_1}{a_1b_0}+1}))$.
Moreover, lower bounds on these ERs are derived as $R_{b0} \geq \log_2(1 + \frac{a_0}{Rb_0\frac{\tau^{2\alpha}}{\alpha+1}+\frac{a_1b_0}{b_1}})$ and $R_{b1} \geq \log_2(1 + \frac{a_1}{Rb_1(\tau^{2\alpha} + \frac{\alpha}{\alpha + 1} \left(1 - \tau^{2(\alpha + 1)/\alpha}\right)))})$.
\end{lemma}
\begin{remark}
Obtaining closed-form expressions for Lemma 4 to 6 is challenging. However, the derived expressions are exact, and the integrals can be easily evaluated numerically in MATLAB since they involve only double integrals without additional nested integrals. Moreover, as illustrated in Fig. 2, \(F_{S_0}(s_0)\) (\(F_{S_1}(s_1)\)) at IBP reaches 1 at approximately 3000 (20) for the corresponding parameters. Consequently, the integral limit at \(R_{a0}\) in Lemma 5 (\(R_{b1}\) in Lemma 6) can be reduced to \(t < \log_2\left(\frac{3000a_0}{Rb_0} + 1\right)\) (or \(t < \log_2\left(\frac{20a_1}{Rb_1} + 1\right)\)). Alternatively, one can approximate the final integral using the incomplete gamma function [\cite{table}, eq. (8-350-1)], leading to a Gauss hyper-geometric function [\cite{table}, eq. (9-100)], based on an infinite summation.
\end{remark}
\begin{remark}\label{ps0smaller}
The probability of the event \( S_0\!\! < S_1 \) is given by:
\[
P(S_0\!\! < \!\!S_1) \!\!= \!\!E_{S_1}[P(S_0 < s_1 \!\!\mid S_1 = s_1)] \overset{(a)}{=} \!\!\!\!\int_{0}^{\infty} \!\!\!\!\!\!F_{S_0}(s_1)f_{S_1}(s_1) ds_1
\]
Where (a) follows from the independence of \( S_0 \) and \( S_1 \). As shown in Fig. \ref{cdf0} (a), \( F_{S_1}(s_1) \) at IBP reaches its maximum value near 20, implying that \( f_{S_1}(s_1) \) is zero beyond 20. In this region, \( F_{S_0}(s_0) \) at IBP is near zero. Thus, we have \( P(S_0 < S_1) \approx 0 \), and Lemma \ref{lemma3new} will rarely occur \footnote{We have derived \( P(S_0 < S_1) \) for different values of \( \lambda_b \) and observed that it is nearly independent of changes in \( \lambda_b \), and When \( N=7 \), \( \tau=0.7 \), \( \alpha=4 \), we have \( P(S_0 < S_1) = 0.0082 \).} Consequently, the ESR of the two users is $
R_{\mathbf{total}} = P(S_0 > S_1)[(R_{a0} - R^0_e) + (R_{a1} - R^1_e)] + P(S_0 < S_1)[(R_{b0} - R^0_e) + (R_{b1} - R^1_e)]
\simeq [(R_{a0} - R^0_e) + (R_{a1} - R^1_e)]$
where \( R^0_e \) is given in (\ref{e55}) and \( R^1_e \) is obtained by replacing \( a_0 \) and \( b_0 \) with \( a_1 \) and \( b_1 \) in (\ref{e55}), respectively.
\end{remark}
Next, we provide a straightforward validation of the theoretical results presented in Lemmas 5 and 6. Note that the ERs are independent of the cell and cluster numbers, $k$ and $\tilde{i}$. Moreover, consider setting \( a_0 = 0 \) in Lemma \ref{lemma2new} and Lemma \ref{lemma3new}. In this case, the ER of the central user becomes 0. This happens because either the limits of the integral become zero, the input of \( F_{S_0}(s_0) \) approaches infinity (making \( F_{S_0} = 1 \) and \( R_{a0} = 0 \)), or setting \( a_0 = 0 \) results in \( \tilde{N}(r_{kk0\tilde{i}}, r_m) = \tilde{M_2}(r_{kk0\tilde{i}}, r_m) = 0 \). This holds regardless of whether the user performs SIC (in Lemma \ref{lemma2new}) or not (in Lemma \ref{lemma3new}). Similarly, setting \( a_1 = 0 \) results in the ER of the second user being zero, even if SIC is performed (as in Lemma \ref{lemma3new}). Moreover, unlike the RP pairing strategy, the ERs of the users in this scenario also depend on the BS density. Moreover, if \(\tau = 0\), the limit of the integral is zero, and \(R_{a0} = R_{b0} = 0\), as expected, since there is no region in the Voronoi cell dedicated to the central user in the IBP strategy. Conversely, if \(\tau = 1\), then \(R_{a1} = R_{b1} = 0\) since there is no region dedicated to the second user. Another insight is that since \( \tilde{N}(r_{kk0\tilde{i}}, r_m) > \tilde{M_2}(r_{kk0\tilde{i}}, r_m) \), it follows that \( R_{a0} > R_{b0} \). Similarly, since \( \tilde{M}(r_{kk1\tilde{i}}, r_m) < \tilde{N_2}(r_{kk1\tilde{i}}, r_m) \), it follows that \( R_{a1} < R_{b1} \). This demonstrates that if a user, either central or secondary, performs SIC, its ER will improve.
\section{Secrecy Outage Probability}\label{SOP}
We derive the SOP for both users in the $\tilde{i}$-th cluster of the $k$-th cell. We define $\tilde{R}_0$ and $\tilde{R}_1$ as the targeted data rates for the central and second users, respectively. The SOP event is defined as: $\mathrm{SOP}=1-P(E_{01}\cap E_{00}\cap E_{11})$, where $\bar{E}_{01}=\lbrace \log(1+\mathrm{SINR}^{w_1}_{k0\tilde{i}})-\log(1+\mathrm{SINR}^{w_1}_{\tilde{e}})<\tilde{R}_1\rbrace$ is the event when the central user cannot decode the second user massage, $w_1$; $\bar{E}_{00}=\lbrace \log(1+\mathrm{SINR}^{w_0}_{k0\tilde{i}})-\log(1+\mathrm{SINR}^{w_0}_{\tilde{e}})<\tilde{R}_0\rbrace,$ is the event when the central user cannot decode its own massage, $w_0$; and $\bar{E}_{11}=\lbrace \log(1+\mathrm{SINR}^{w_1}_{k1\tilde{i}})-\log(1+\mathrm{SINR}^{w_1}_{\tilde{e}})<\tilde{R}_1\rbrace.$ is the event when the second user cannot decode its own massage, $w_1$.
\subsection{RP strategy} According to (\ref{hello22}), (\ref{hello2}) and due to our assumption on user ordered based on $S_i$ defined in Section \ref{model}: if $1+\mathrm{SINR}^{w_1}_{k1\tilde{i}}>2^{\tilde{R}_1}(1+\mathrm{SINR}^{w_1}_{\tilde{e}})$, then $1+\mathrm{SINR}^{w_1}_{k0\tilde{i}}>2^{\tilde{R}_1}(1+\mathrm{SINR}^{w_1}_{\tilde{e}})$, which means ${E}_{01}\subseteq {E}_{11}$. Thus, the SOP is equal to $\mathbf{SOP}^{\mathbf{t}}_{\mathbf{RP}}=1-(1-P^{w_0}_{\mathrm{out}})(1-P^{w_1}_{\mathrm{out}})$,
where $P^{w_0}_{\mathrm{out}}$ and $P^{w_1}_{\mathrm{out}}$ stands for $P(\bar{E}_{00})$ and $P(\bar{E}_{11})$, the SOPs of the central user and the second user, respectively. Lemma~\ref{lemma4} (proof at Appendix \ref{lemma4p}) derives these probabilities using two distinct methods.
\begin{lemma}\label{lemma4}
The SOPs of the central and the second user are denoted by $P^{w_0}_{\mathrm{out}}$ and $P^{w_1}_{\mathrm{out}}$ respectively. If $0>(\frac{a_1b_0}{a_0b_1}+1)2^{-\tilde{R_1}}-1$, $P^{w_1}_{\mathrm{out}}=1$, otherwise we have:

Method 1:
\begin{align}
&P^{w_0}_{\mathrm{out}}=\int_{0}^{\infty}
(\sum_{n=0}^{N}(-1)^{n}{N\choose{n}}\frac{-\alpha+1}{-\alpha+1-\eta n(\frac{Rb_0(2^{\tilde{R}_0}(1+z)-1)}{a_0})})^2\nonumber\\
&P^{w_1}_{\mathrm{out}}=\!\!\!\!\int_{(\frac{a_1b_0}{a_0b_1}+1)2^{-\tilde{R_1}}-1}^{\infty}\!\!\!\!\!\!\!\!\!\!\!\!\!\!\!\!\!\!\!\!\!\!\!\!\!\!\!\!\!\!f_{\mathrm{SINR}^{w_1}_{\tilde{e}}}(z)dz,+\int_{0}^{(\frac{a_1b_0}{a_0b_1}+1)2^{-\tilde{R_1}}-1}\!\!\!\!\!\!\!\!\!\!\!\!\!\!\!\!\!\!\!\!\!\!\!\!
(1-(\sum_{n=1}^{N}(-1)^{n+1}{N\choose{n}}\nonumber\\&\frac{-\alpha+1}{-\alpha+1-\eta n(\frac{(2^{\tilde{R}_1}(1+z)-1)Rb_1}{a_1-\frac{(2^{\tilde{R}_1}(1+z)-1)a_0b_1}{b_0}})
})^2)f_{\mathrm{SINR}^{w_1}_{\tilde{e}}}(z)dz\label{pout13}
\end{align}

Method 2:
\begin{align}
&P^{w_0}_{\mathrm{out}}=\int_{0}^{\infty}e^{2(1-\alpha)\frac{a_0}{Rb_0(2^{\tilde{R}_0}(1+z)-1)}}f_{\mathrm{SINR}^{w_0}_{\tilde{e}}}(z)dz\nonumber\\
&P^{w_1}_{\mathrm{out}}=\int_{0}^{(\frac{a_1b_0}{a_0b_1}+1)2^{-\tilde{R_1}}-1}\!\!\!\!\!\!\!\!\!\!\!\!\!\!\!\!\!\!\!\!\!\!\! [(1-(1-e^{(1-\alpha)\frac{a_1-\frac{(2^{\tilde{R}_1}(1+z)-1)a_0b_1}{b_0}}{(2^{\tilde{R}_1}(1+z)-1)Rb_1}}\!\!\!)^2)\nonumber\\&
f_{\mathrm{SINR}^{w_1}_{\tilde{e}}}(z)dz]
+\!\!\!\int_{(\frac{a_1b_0}{a_0b_1}+1)2^{-\tilde{R_1}}-1}^{\infty}f_{\mathrm{SINR}^{w_1}_{\tilde{e}}}(z)dz
\end{align}
where the closed-form of $f_{\mathrm{SINR}^{w_0}_{\tilde{e}}}$ and $f_{\mathrm{SINR}^{w_1}_{\tilde{e}}}$ are given at Appendix \ref{lemma4p}.
\end{lemma}
\subsection{IBP strategy}\label{IBPstrategy}
The total SOP of the central and second users, $\mathbf{SOP}^{\mathbf{total}}_{\mathbf{out}}$, is defined as:
$\mathbf{SOP}^{\mathbf{t}}_{\mathbf{IBP}}\!=\!P(S_0\!>\!S_1)[1-(\!1\!-\!P^{w_0}_{\mathrm{aout}})(1\!-\!P^{w_1}_{\mathrm{aout}})]
\!+\!P(S_0\!<\!S_1)[1-\!(1\!-\!\!P^{w_0}_{\mathrm{bout}})(1\!-\!P^{w_1}_{\mathrm{bout}})]\overset{(a)} \simeq [1-(\!1\!-\!P^{w_0}_{\mathrm{aout}})(1\!-\!P^{w_1}_{\mathrm{aout}})]$
Where (a) is due to Remark \ref{ps0smaller}. \( P^{w_0}_{\mathrm{aout}} \) and \( P^{w_1}_{\mathrm{aout}} \) represent the SOPs of the central user and the second user, respectively, when \(S_0>S_1 \) and \( P^{w_0}_{\mathrm{bout}} \) and \( P^{w_1}_{\mathrm{bout}} \) denote the SOPs of the central user and the second user, respectively, when \( S_0<S1\). By following the same approach as in the proof of Lemma \ref{lemma4} and utilizing equations (\ref{laplas2new}) (c), (\ref{laplas3}) (c), (\ref{cdfs0new}), and (\ref{cdfs1new}), along with the closed forms of \( f_{\mathrm{SINR}^{w_0}_{\tilde{e}}} \) and \( f_{\mathrm{SINR}^{w_1}_{\tilde{e}}} \) given in Appendix \ref{lemma4p}, we can derive \( P^{w_0}_{\mathrm{aout}} \), \( P^{w_1}_{\mathrm{aout}} \), \( P^{w_0}_{\mathrm{bout}} \), and \( P^{w_1}_{\mathrm{bout}} \). Due to space limitations, we do not provide their expressions here.
\begin{remark}
As illustrated in Fig. 2 (b), \( f_{\mathrm{SINR}^{w_0}_{\tilde{e}}} \) and \( f_{\mathrm{SINR}^{w_1}_{\tilde{e}}} \) reach 0 at approximately 0.5 for the corresponding parameters. Consequently, the infinity limits of the integral at \( P^{w_0}_{\mathrm{out}} \) and \( P^{w_1}_{\mathrm{out}} \) in Lemma 7 are reduced to 0.5. Thus, by plotting the closed-form expression of \( f_{\mathrm{SINR}^{w_0}_{\tilde{e}}} \) and \( f_{\mathrm{SINR}^{w_1}_{\tilde{e}}} \) as derived in Appendix G, we can reduce the \(\infty\) at the outermost integral in Lemma 7 to the value at which \( f_{\mathrm{SINR}^{w_0}_{\tilde{e}}} \) (or \( f_{\mathrm{SINR}^{w_1}_{\tilde{e}}} \)) is zero, which depends on the system parameters. We note that this value will never be infinity as the maximum value of \( \mathrm{SINR}^{w_0}_{\tilde{e}} \) (or \( \mathrm{SINR}^{w_1}_{\tilde{e}} \)) is \( -\log_2((\frac{Rb_0}{d_ia_0})-1) \) (or \( -\log_2((\frac{Rb_1}{d_ia_1})-1) \)).
\end{remark}
To gain further insights, consider setting \(a_0 = 0\) in \(P^{w_0}_{\mathrm{out}}\) given in Lemma \ref{lemma4}. Since 
$\lim_{a_0 \rightarrow 0} (\sum_{n=0}^{N} (-1)^n {N \choose n} \frac{-\alpha + 1}{-\alpha + 1 - \eta n \infty})=1$, we have \(P^{w_0}_{\mathrm{out}} = \int_{0}^{\infty} f_{\mathrm{SINR}^{w_0}_{\tilde{e}}}(z) \, dz = 1\). Similarly, by setting \(a_1 = 0\), we have \(P^{w_1}_{\mathrm{out}} = 1\). These results are as expected since the users receive no signal power.
\begin{remark}:Although the IBP strategy appears to outperform RP in most scenarios (as shown in Section \ref{simulations}), RP exhibits lower computational complexity in deriving performance metrics. In addition, under IBP, the central user may fail to perform SIC in certain cases (Section \ref{newlabeladd}), necessitating separate derivations for ergodic rate and outage probability despite the low probability of this event (Remark \ref{ps0smaller}). Moreover, IBP reduces the need for precise relative distance knowledge compared to RP, as it classifies users based on partitioned Poisson-Voronoi (PV) regions rather than normalized distances.
\end{remark}
\begin{remark}: For a total of $N$ users (with $N$ odd), there are $I = \frac{N + 1}{2}$ clusters—specifically, $\frac{N - 1}{2}$ clusters each containing two NOMA users, and one cluster with the single user receiving full power. This approach ensures that the NOMA users do not experience any rate reduction, as they continue to share resources via power-domain multiplexing.
Consequently, the ESR and SOP for the paired NOMA users follow Lemmas 2–3 and 5–7, while the ESR and SOP for the unpaired user are derived similarly but with full power allocation . The overall ESR and SOP for a randomly selected user are computed as weighted averages, where the weights $\frac{N-1}{N}$ and $\frac{1}{N}$ correspond to the probabilities of selecting a NOMA user or a singleton user in a cluster, respectively.\end{remark}
\section{Simulation and Numerical Results}\label{simulations}
In this section, the analytical results from Sections \ref{ergodic_rate} and \ref{SOP} are illustrated through numerical plots. The SER and SOP for the RP and IBP strategies are evaluated as functions of system parameters. Additionally, results are compared to a massive MIMO-OMA case, specifically time-division multiple access (TDMA). The SINR for an OMA user is derived similarly to (\ref{hello}), by setting \(a_1 = 0\), \(b_1 = 0\), \(a_0 = 1\), and \(b_0 = 1\), allocating all power to one user. The ER for the OMA user is derived similarly to (\ref{R_Df}), considering a factor of \(\frac{1}{2}\). Leakage to the strongest eavesdropper and SOP in OMA is obtained similarly to (\ref{e55}) and (\ref{pout13}) by applying the aforementioned changes. Simulation results are based on 4000 (in some cases, 5000) randomly seeded channel and node location realizations. In each iteration, a Poisson point process (PPP) with density \(\lambda_b\) is generated in a 4 km x 4 km area for BS locations. Each cell is assumed to have five clusters (\(I = 5\)). For user locations, a PPP with a high density (\(120\lambda_b\)) is generated. Each user is assigned to its nearest BS, and each BS selects \(2I\) of its assigned users to serve, ignoring the others. For attackers, a PPP with density \(\lambda_e\) is generated. Analytical results converge to the Monte Carlo simulations with a small value of the shape parameter of the Gamma random variable (\(N = U = 7\)) in most evaluated cases, as plotted in the results (defined in Appendix \eqref{lemma2p})\footnote{Larger \(U\) or \(N\) values do not change the results.}. The parameters used are \(I=5\), \(\alpha=4\), \(b_0=0.4\), \(b_1=0.6\), \(a_0=0.1\), \(a_1=0.9\), \(d_{\tilde{i}}=0.2\), and \(\tau=0.7\), unless otherwise stated. In all figures, \(M_1\) and \(M_2\) denote the numerical results for the first and second method derivations. Both methods closely match in all figures, validating each other's accuracy.

\begin{figure*}
    \centering
    \subfigure[Random-pairing (RP)]{\includegraphics[scale=.30]{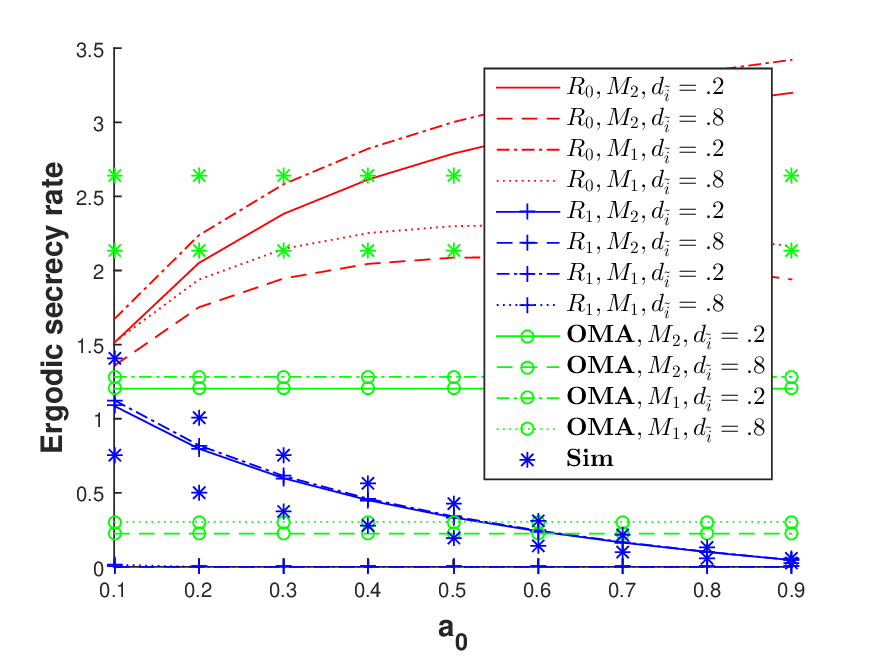}}\label{a}
    \subfigure[Interference-based-pairing (IBP)]{\includegraphics[scale=.30]{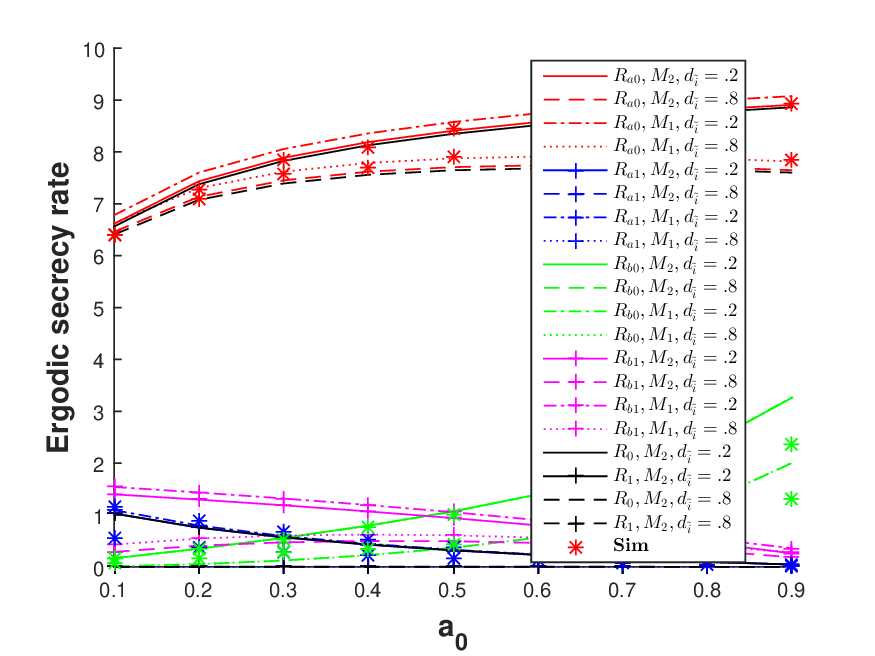}}\label{b}
    \subfigure[Eavesdropper]{\includegraphics[scale=.30]{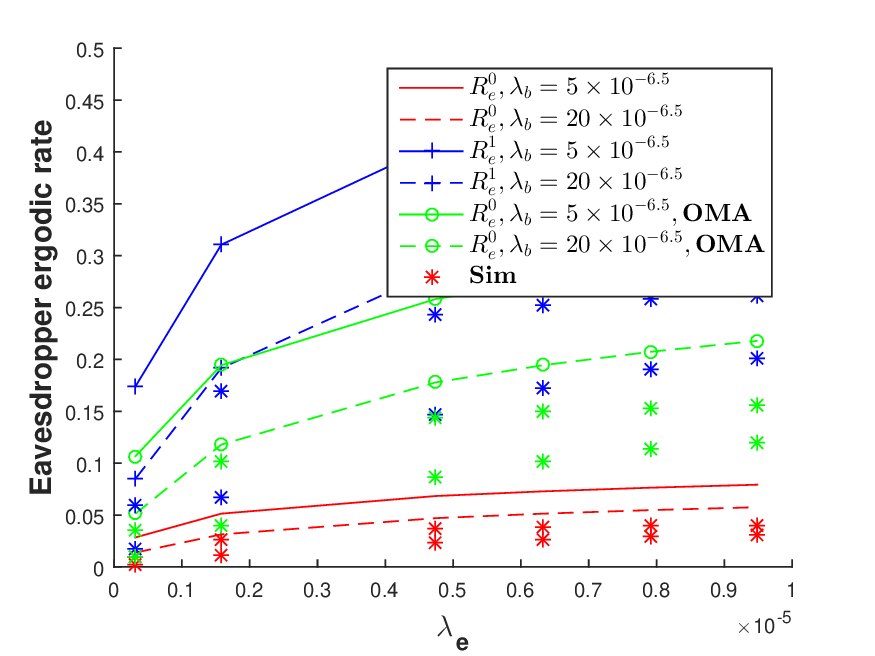}}\label{c}
    \subfigure[Interference-based-pairing (IBP)]{\includegraphics[scale=.30]{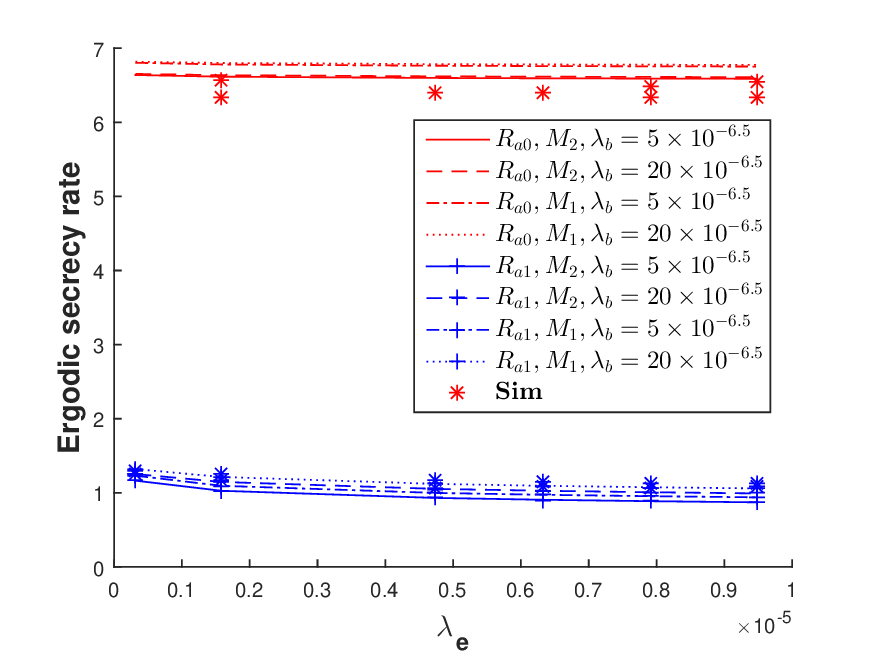}}\label{d}
    \caption{Ergodic secrecy rate (ESR)}
    \label{ratea0di}
\end{figure*}
Fig. \ref{ratea0di} (a)-(b) show the ESR for RP and IBP, respectively, versus the power coefficient of the central user $(a_0)$, for two different power coefficients of the eaves assigned to users in cluster \(\tilde{i}\) (\(d_{\tilde{i}}\)), when $\lambda_b=\lambda_e=15\times 10^{-6.5}$. Notably, in Fig. \ref{ratea0di} (b), the curves for \( R_{b0}, M_2, d_{\tilde{i}} = 0.2 \) and \( R_{b0}, M_2, d_{\tilde{i}} = 0.8 \) perfectly match those for \( R_{b0}, M_1, d_{\tilde{i}} = 0.2 \) and \( R_{b0}, M_1, d_{\tilde{i}} = 0.8 \), respectively. Also, \( R_0 = P(S_0 > S_1) R_{a0} + P(S_0 < S_1) R_{b0} \) and \( R_1 = P(S_0 > S_1) R_{a1} + P(S_0 < S_1) R_{b1} \) are the total ESRs of the central and second user in IBP, respectively. As observed, \( R_0 \approx R_{a0} \) and \( R_1 \approx R_{a1} \), validating the result in Remark \ref{ps0smaller} \footnote{In all the other figures, we show only the case when \( S_0 > S_1 \) (\( R_{a0} \) and \( R_{a1} \) or \( P^{w_0}_{\mathrm{aout}} \) and \( P^{w_1}_{\mathrm{aout}} \)), as the total performance in IBP is almost entirely determined by this case.}. Throughout, increasing \(a_0\) results in an increase (decrease) in the central (second) user's ESR. This means that although the signal strength to both the eaves and central (second) user increases (decrease), the signal power to the user experiences a higher increment (decrement), leading to an increase (decrease) in the user's ESR. Moreover, increasing \(d_i\) decreases the ESR for both users, especially affecting the second user as it leads to the second user's ESR becoming 0, except in the case of \( R_{b1} \). This is because, in this case, the second user performs SIC and has a higher rate compared to the second user in RP or IBP. Moreover, the ESR for the central user is higher than that for the second user (since it performs SIC, while the second user suffers from central user interference), except for \( R_{b0} \) and \( R_{b1} \) when \( a_0 < 0.5 \). Furthermore, the ESR of the OMA system is lower than that of the central user in RP and IBP but higher than that of the second user. In summary, the overall ESR of the system improves with the utilization of NOMA. Also, the ESR of the central user in the IBP is higher than that for the RP. However, the ESR of the second user in the IBP is a bit lower than that for the RP. This is because the IBP forces the central and second users to be in particular regions, and the second user cannot be in the area where the central user is located. Therefore, in the RP (IBP), the second (central) user can be closer to the BS that serves it compared to the second (central) user in the IBP (RP). Another observation is that for the central user, with increasing \(a_0\), the gap between the ESR when \(d_i = 0.2\) and \(d_i = 0.8\) increases, meaning that increasing the power coefficient of eaves when \(a_0\) is high has a more destructive effect on the ESR. Moreover, increasing \(d_i\) has a more destructive impact on the second user compared to decreasing \(a_1\) (increasing \(a_0\)). For example, in Fig. \ref{ratea0di} (a), by increasing \(d_i\) by 0.6, the ESR of the second user is approximately 0 across the entire range of \(a_0\). However, at \(d_i = 0.2\), increasing \(a_0\) by 0.9 results in the ESR of the second user becoming 0.
\begin{figure}
\centering
    \subfigure[RP and Eavesdropper]{\includegraphics[scale=.29]{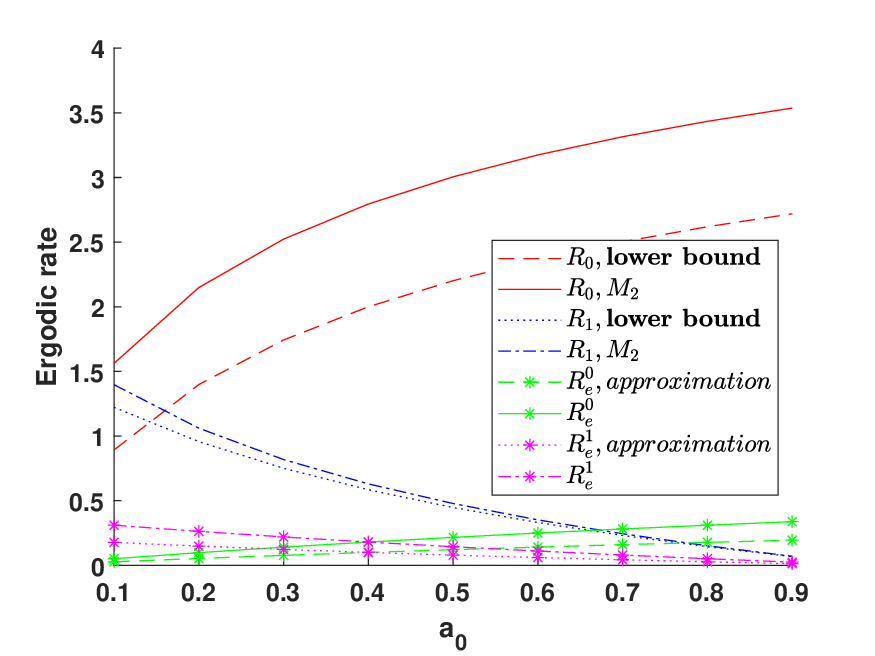}}\label{a}
    \subfigure[IBP]{\includegraphics[scale=.29]{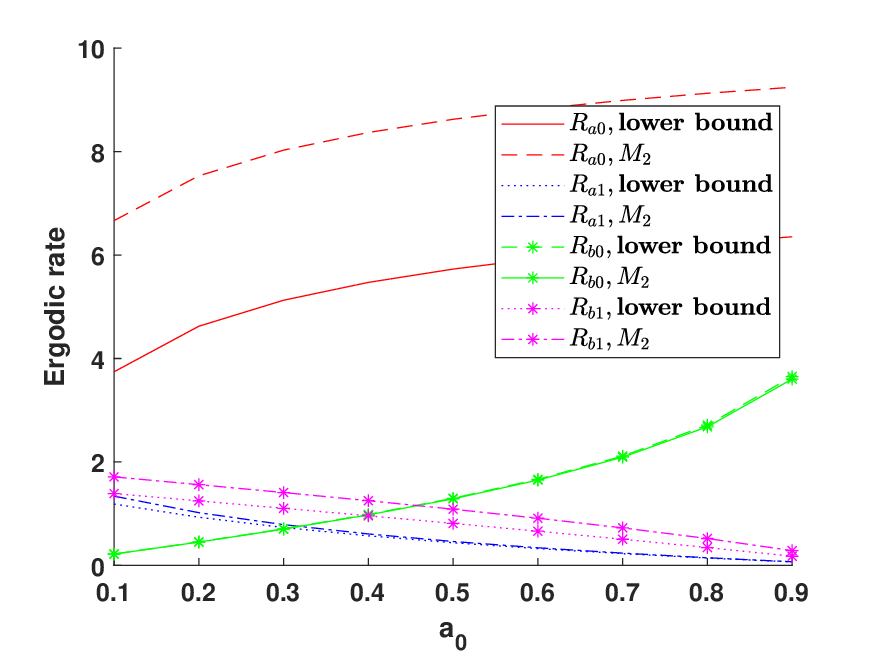}}\label{b}
    \caption{Lower bound on ERs }
    \label{lowerboundERs}
\end{figure}
Fig. \ref{ratea0di} (c), and (d), show the upper bound of the rate of the eavesdropper, and ESR at IBP, respectively, versus the attacker's density for two different BS densities. \footnote{We note that as shown in Lemma \ref{lemma2}, the ER of the users in the RP strategy is independent of \(\lambda_b\). Thus, any change in the ESR is due to the change in the eavesdropper rate. Therefore, instead of plotting the ESR for the RP strategy versus \(\lambda_e\) and \(\lambda_b\), we have plotted the eavesdropper rate.} In Fig. \ref{ratea0di} (c), as \(\lambda_e\) increases, the eavesdropper ER increases due to improved attacker SINR, while increasing \(\lambda_b\) leads to decreased ER for the most harmful eavesdropper across all \(\lambda_e\) ranges. This occurs because closer proximity between eavesdroppers and BSs boosts desired signal power for attackers but also increases interference from other BSs. However, the decreasing factor outweighs the increasing one, resulting in an overall decrease in eavesdropper ER. Thus, in Fig. \ref{ratea0di} (d), we see a slight increase in the ESR. By comparing Fig. \ref{ratea0di} (c,d) and Fig. \ref{ratea0di} (a,b), we observe that the ESR is more sensitive to \(a_0\) than to the eavesdropper or BS density, as the slopes of the curves are steeper.
\begin{figure}
\centering
    \subfigure[$\lambda_b=\lambda_e=5\times 10^{-6.5}$]{\includegraphics[scale=.29]{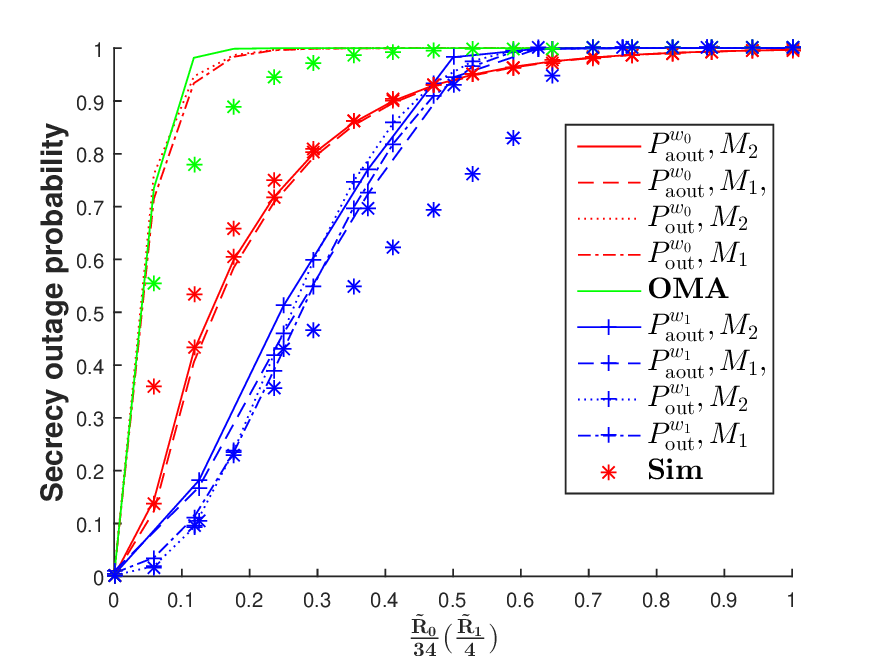}}\label{a}
    \subfigure[$\lambda_b=\lambda_e=15\times 10^{-6.5}$]{\includegraphics[scale=.29]{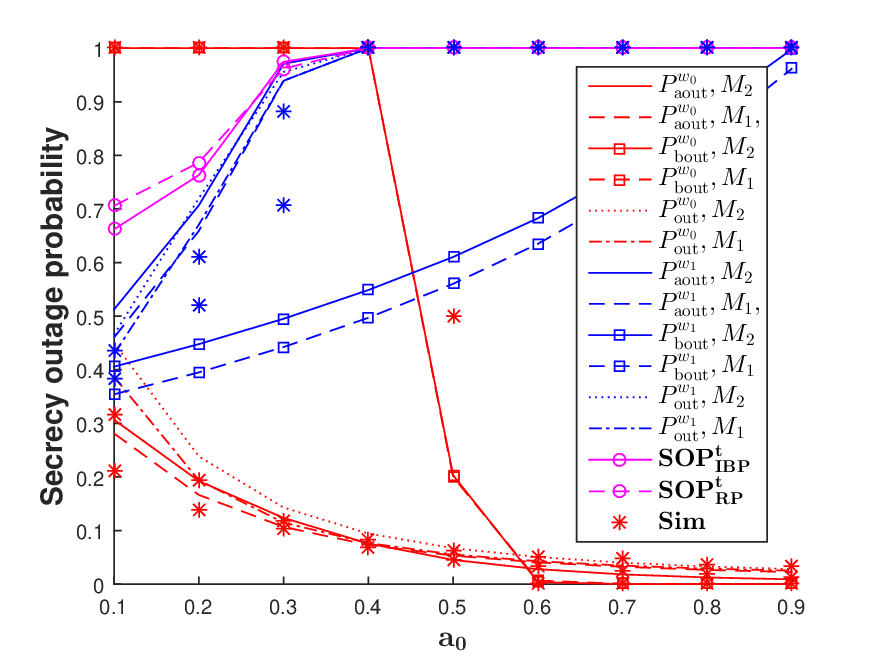}}\label{b}
    \caption{Secrecy outage probability (SOP) versus $\tilde{R}_{0}/34$ ($\tilde{R}_{1}/4$), and $a_0$. }
    \label{outagea0tresh}
\end{figure}
Fig. \ref{lowerboundERs} (a) and (b) show the lower bounds on the ergodic rates of the users under RP and IBP, as well as an approximation of the eavesdropper's ergodic rate, plotted against $a_0$. We observe that the lower bounds on the second user's ergodic rates in both RP and IBP—whether it performs SIC or not—are tight. Additionally, the lower bound for the central user in IBP, when it does not perform SIC, is also tight.

Fig. \ref{outagea0tresh} (a) shows the SOP of the users in RP, IBP, and OMA versus scaled target rates. Among the central users, the central user in the IBP strategy outperforms all others. Additionally, the SOP of OMA is worse than that of the central user in both the RP and IBP strategies. From the slope of the curve, we observe that the SOP of the central user in IBP is less sensitive to the target rate and reaches 1 at about \(\tilde{R}_0 = 34\) bits per channel use (BPCU), while the SOP for the others reaches 1 at about \(\tilde{R}_0 = 10\) for OMA and the central user in RP, and \(\tilde{R}_1 = 4\) for the second users. At lower target rates, the SOP of the second user in RP outperforms that of the second user in IBP; however, at higher target rates, the reverse is true. Moreover, there is a tolerable threshold for the target rate above which the attack will be successful, where the SOP becomes high.
Fig. \ref{outagea0tresh} (b) shows the SOP versus \(a_0\) for users in RP and IBP. The targeted ESR are \(\tilde{R}_0 = 1\) (for IBP when it performs SIC, \(\tilde{R}_0 = 3\)), and \(\tilde{R}_1 = 1\) BPCU. As \(a_0\) increases, the SOP of the central user decreases and the SOP of the second user increases in all scenarios. However, the total SOP in both RP and IBP increases, meaning the secondary user affects the SOP more significantly with changes in \(a_0\). Moreover, the total SOP in RP is higher than that in IBP until \(a_0 < 0.5\); beyond this range, the total SOP is 1 in both cases \footnote{We note that the total SOP is high in both scenarios due to the selected target rate, and by changing these targets, one can achieve a smaller SOP as shown in Fig. \ref{outagea0tresh} (a).}.
\section{Conclusion}\label{conclusion}
We analyzed the performance of physical layer security techniques for a random massive MIMO-NOMA network in the presence of active eavesdroppers in two-user pairing strategies. We characterized the ESR and SOP, and by using numerical and simulation results, we show that NOMA can improve system performance in comparison with OMA. Also, when using NOMA, one of the pairing strategies improved system performance.
\bibliographystyle{IEEEtran}
\bibliography{ref}
\appendices
\section{Proof of lemma \ref{lemma1}} \label{lemma1p}
As the intended signal for the user in the $\tilde{i}$-th cluster of the $k$-th cell is $ s_{k0\tilde{i}} $, by defining $\mathcal{D}=\lbrace (i,m)\vert i\in\lbrace 1,...,I\rbrace , m\in \lbrace 0,1\rbrace\rbrace$, we have $y_{k0\tilde{i}}=I_0+I_1+I_2+w_d$. $I_0$, $I_1$, and $I_2$, respectively, are the desired signal for the central user, the interference caused by the second user in the same cell, and the interference caused by the users in other cells, respectively. Moreover, $\mathrm{SINR}^{w_0}_{k0\tilde{i}}=\frac{E[I^2_0]}{E[y_{k0\tilde{i}}]}$. According to (\ref{law1}) and (\ref{law2}), in $I_0$ only $\mathbf{h}^H_{kk0\tilde{i}}\mathbf{h}_{kk0\tilde{i}}$ is proportional to $M$ and the other terms tend to zero. In $I_1$, all terms tend to zero (the term including $ \sqrt {a_1}s_{k1 \tilde{i}} $ will be zero because of SIC at the central user), and only $\mathbf{h}^H_{lk0\tilde{i}}\mathbf{h}_{lk0\tilde{i}}$ will be a non-vanishing term. Therefore, for $M \rightarrow \infty$ (massive MIMO regime), we have:
\begin{align}
\mathrm{SINR}^{w_0}_{k0\tilde{i}}=\frac{\frac{MP_da_0r^{-2\alpha}_{kk0\tilde{i}}}{2}}{\sigma^2_n+\frac{MP_db_0}{2}\sum_{l\neq k}^{}r^{-2\alpha}_{lk0\tilde{i}}\big(\sum_{m=0}^{1}\frac{a_m}{b_m}\big)}.\label{conclu0}
\end{align}
In addition, when $M \rightarrow \infty$, $\sigma^2_n$ is negligible in comparison to other terms; thus, by defining $ R\triangleq\sum_{m=0}^{1}\frac{a_m}{b_m} $ we obtain $\mathrm{SINR}^{w_0}_{k0\tilde{i}}=\frac{a_0r^{-2\alpha}_{kk0\tilde{i}}}{Rb_0\sum_{l\neq k}^{\infty}r^{-2\alpha}_{lk0\tilde{i}}}$. Following similar steps, $ \mathrm{SINR}^{w_1}_{k0\tilde{i}} $ and $\mathrm{SINR}^{w_1}_{k1\tilde{i}}$ in the massive MIMO regimes are expressed \footnote{the derived SINR are nearly exact due to the properties of massive MIMO, which state that the variance of signal power, proportional to \(|\mathbf{h}^H_{klmi}\mathbf{h}_{\hat{k} \hat{l} \hat{m} \hat{i}}|^2\) and \(|\mathbf{g}^H_{ke}\mathbf{g}_{\hat{k} \tilde{e}}|^2\), decreases with \(M^2\), as discussed in \cite{ngo2014aspects}.}. This completes the proof.
\section{Proof of Lemma \ref{lemma2}}\label{lemma2p}
First, we show that the central user can perform SIC. To show this, we must check if $ \mathrm{SINR}^{w_1}_{k0\tilde{i}} > \mathrm{SINR}^{w_1}_{k1\tilde{i}} $. Using (\ref{hello22}) and (\ref{hello2}) and after some simple mathematical manipulations, we obtain $ \frac{r^{-2\alpha}_{kk0\tilde{i}}}{\sum_{l\neq k}^{}r^{-2\alpha}_{lk0\tilde{i}}}>\frac{r^{-2\alpha}_{kk1\tilde{i}}}{\sum_{l\neq k}^{}r^{-2\alpha}_{lk1\tilde{i}}}$ which holds by the definition of central and second users mentioned in section \ref{orderingdef}. Thus, the central user can perform SIC.
In the following, we provide two distinct approaches for deriving the ERs of the central and second users. Then, we derive lower bounds of these ERs.

\textbf{Method 1:} By defining $S_0=\frac{r_{kk0\tilde{i}}^{-2\alpha}}{\sum_{l \neq k}^{}r_{lk0\tilde{i}}^{-2\alpha}}$ and $2^t-1=T$, we have:
\begin{align}
R_0&=E[\log(1+\mathrm{SINR}^{w_0}_{k0\tilde{i}})]\!\!=\!\!\int_{0}^{\infty}\!\!\!P(\log( 1+\mathrm{SINR}^{w_0}_{k0\tilde{i}})>t)dt\nonumber\\
&=\!\!\int_{0}^{\infty}\!\!(1-F_{S_0}(\frac{TRb_0}{a_0}))dt.\label{R_d}
\end{align}
Moreover, by defining $S_1=\frac{r_{kk1\tilde{i}}^{-2\alpha}}{\sum_{l\neq k}^{}r_{lk1\tilde{i}}^{-2\alpha}}$ and using the same approach as (\ref{R_d}), the ER of the second user is:
\begin{align}
R_1&=\int_{0}^{\log{\frac{a_1b_0}{a_0b_1}+1}}(1-F_{S_1}(\frac{TRb_1}{a_1-\frac{Ta_0b_1}{b_0}}))dt,\label{R_d2}
\end{align}
where $(a)$ is due to:$\displaystyle{\lim_{\sum_{l\neq k}^{}r_{lk1\tilde{i}}^{-2\alpha}\to 0} (\mathrm{SINR}^{w_1}_{k1\tilde{i}})=\frac{a_1b_0}{a_0b_1}}$. Thus, we need to derive the CDF of $S=\frac{r_{kki\tilde{i}}^{-2\alpha}}{\sum_{l\neq k}^{}r_{lki\tilde{i}}^{-2\alpha}}$, where $r_{kki\tilde{i}}$ and $r_{lki\tilde{i}}$ are distances between a randomly selected user in $\tilde{i}$-th cluster of the $k$-th cell and its serving BS and interfering BSs, respectively. We use order statistics to compute the CDF of $S_0=\max(S_0,S_1)$ and $S_1=\min(S_0,S_1)$ based on CDF of $S$, as follows:
\begin{align}
F_{S_1}(s_1)=2F_{S}(s_1)-F^2_{S}(s_1)\quad \quad,  F_{S_0}(s_0)&=F^2_{S}(s_0).\label{F_S}
\end{align}
The CDF of $S$ is expressed as:
\begin{align}
F_{S}(s)
&=1-\int_{0}^{\infty}P(\frac{r_{kki\tilde{i}}^{-2\alpha}}{\sum_{l\neq k}^{}r_{lki\tilde{i}}^{-2\alpha}}>s|r_{kki\tilde{i}}=r)f_{r_{kki\tilde{i}}}(r)dr,\label{F_srevise}
\end{align}
As each user connects to its nearest BS, no BS must be in a disk of radius $r$. we have:
\begin{align}
F_{r_{kki\tilde{i}}}(r)=1-e^{-\lambda_b \pi r^2}, \quad
f_{r_{kki\tilde{i}}}(r)=2\pi \lambda_b r e^{-\lambda_b \pi r^2}.\label{pdf}
\end{align}
Therefore, we have:
\begin{align}
&P(\frac{r_{kki\tilde{i}}^{-2\alpha}}{\sum_{l\neq k}^{}r_{lki\tilde{i}}^{-2\alpha}}>s|r_{kki\tilde{i}}=r)=P(1>s(\frac{\sum_{l\neq  k}^{}r_{lki\tilde{i}}^{-2\alpha}}{r^{-2\alpha}}))\!\nonumber\\
&\overset{(a)}{\simeq}\!P(g>s(\frac{\sum_{l\neq  k}^{}r_{lki\tilde{i}}^{-2\alpha}}{r^{-2\alpha}}))\!\overset{(b)}
{=}\!P(g>sA)=1-E_A[F_g(sA)]\nonumber\\
&\overset{(c)}=\sum_{n=1}^{N}(-1)^{n+1}{N\choose{n}}E[e^{-sn\eta A}],\label{ea}
\end{align}
(a) follows from the fact that we approximate 1 with a dummy Gamma random variable, $g$, with unit mean and the shape parameter of $N$ such that \cite{Dohe2002math}:$\displaystyle{\lim_{N \to \infty}\frac{N^Ng^{N-1}e^{-Ng}}{\Gamma(N)}=\delta(g-1)}$, where $\delta(.)$ is the Dirac delta function and $\Gamma(.)$ is the Gamma function defined as $\Gamma(a)=\int_{0}^{\infty}e^{-t}t^{a-1}dt$. (b) follows from defining
$A\triangleq \frac{\sum_{l\neq k}^{}r_{lki\tilde{i}}^{-2\alpha}}{r^{-2\alpha}}$. (c) follows from the fact that we use the Alzer inequality in \cite[Appendix A]{Bai2015alzer} to give a tight approximation of the CDF, as $F_g(A)=(1-e^{-AN(N!)^\frac{1}{N}})^N$; then we use a binary extension, defining $\eta=N(N!)^\frac{-1}{N} $. Moreover, we have:
\begin{align}
E[e^{-sn\eta A}]\overset{(a)}{=}e^{\int_{r}^{}[\exp(-\eta nsx^{-2\alpha}r^{2\alpha})-1]2\pi \lambda_b xdx},\label{A}
\end{align}
where (a) is due to the probability generating functional (PGFL) of the PPP \cite{book} and the fact that distances from other BSs to the user must be larger than $ r_{kki\tilde{i}}$. So, by combining (\ref{ea}) and (\ref{A}), and replacing the result and (\ref{pdf}) into (\ref{F_srevise}), the CDF of $S$ is:
\begin{align}
&F_{S}(s)
\overset{(a)}=1-\int_{0}^{\infty}\!\!\sum_{n=1}^{N}(-1)^{n+1}{N\choose{n}}e^{\frac{\pi\lambda_b s\eta nr^2}{-\alpha+1}}
2\pi \lambda_b r e^{-\lambda_b \pi r^2}\!\!\!dr\nonumber\\&
\overset{(b)}=1-\sum_{n=1}^{N}(-1)^{n+1}{N\choose{n}}\frac{-\alpha+1}{-\alpha+1-s\eta n}\label{F_Sfinal}
\end{align}
Where (a) arises from the first-order Taylor approximation, $e^{a} \approx 1 + a$. This approximation holds well when $a$ is close to zero (Taylor series). Since $\alpha>0$, the term $-s\eta nr^{2\alpha}x^{-2\alpha}$ tends to zero as $x$ increases. Therefore, we can employ the mentioned approximation to derive (a). For (b), we use $\int e^{Ar^2}rdr = \frac{e^{Ar^2}}{2A}$.
Finally, by substituting (\ref{F_Sfinal}) into (\ref{F_S}), and using (\ref{R_d}) and (\ref{R_d2}), $R_0$ and $R_1$ are derived as Lemma \ref{lemma2}.

\textbf{Method 2:} By defining $\tilde{S_0}=\frac{\sum_{l \neq k}^{}r_{lk0\tilde{i}}^{-2\alpha}}{r_{kk0\tilde{i}}^{-2\alpha}}$, $2^t-1=T$, and following the same approach as (\ref{R_d}) we have $R_0=\!\!\int_{0}^{\infty}\!\!F_{\tilde{S_0}}(\frac{a_0}{TRb_0})dt$. Moreover, by defining $\tilde{S_1}=\frac{\sum_{l\neq k}^{}r_{lk1\tilde{i}}^{-2\alpha}}{r_{kk1\tilde{i}}^{-2\alpha}}$, the ER of the second user is $R_1=\int_{0}^{\log{\frac{a_1b_0}{a_0b_1}+1}}F_{\tilde{S_1}}(\frac{a_1-\frac{Ta_0b_1}{b_0}}{TRb_1})dt$. Thus, we need to derive the CDF of $\tilde{S}=\frac{\sum_{l\neq k}^{}r_{lki\tilde{i}}^{-2\alpha}}{r_{kki\tilde{i}}^{-2\alpha}}$ and then use order statistics to compute the CDF of $\tilde{S_0}=\min(\tilde{S_0},\tilde{S_1})$ and $\tilde{S_1}=\max(\tilde{S_0},\tilde{S_1})$ as follows:
\begin{align}
F_{\tilde{S_1}}(\tilde{s_1})=F^2_{\tilde{S}}(\tilde{s}),\quad \quad F_{\tilde{S_0}}(\tilde{s_0})&=2F_{\tilde{S}}(\tilde{s})-F^2_{\tilde{S}}(\tilde{s}).\label{F_Snew}
\end{align}
The CDF of $\tilde{S}$ is expressed as:
\begin{align}
F_{\tilde{S}}(\tilde{s})=L^{-1}\{\frac{1}{s}L\{\tilde{S}\}\}\overset{(a)}=L^{-1}\{\frac{1}{s}E[e^{-s\tilde{S}}]\},\label{laplas}
\end{align}
where \( L\{.\} \) and \( L^{-1}\{.\} \) are the Laplace and inverse Laplace transforms, respectively. (a) follows from the definition of the Laplace transform. Moreover, we have:
\begin{align}
&E[e^{-s\tilde{S}}]\overset{(a)}=E\{E[e^{-sr^{2\alpha}\sum_{l\neq k}^{}r_{lki\tilde{i}}^{-2\alpha}}|r_{kki\tilde{i}}=r]\}\nonumber\\
&\overset{(b)}{=}E[e^{\int_{r}^{}[\exp(-sx^{-2\alpha}r^{2\alpha})-1]2\pi \lambda_b xdx}|r_{kki\tilde{i}}=r]\nonumber\\
&\overset{(c)}{=}\int_{0}^{\infty}e^{\frac{2\lambda_b \pi s r^2}{-2\alpha+2}}2\pi\lambda_b r e^{-\lambda_b \pi r^2}\overset{(e)}{=}\frac{\alpha-1}{s+\alpha-1}\label{Anew}
\end{align}
where (a) is due to the law of total expectation, (b) is due to the PGFL of the PPP \cite{book} and the fact that distances from other BSs to the user must be larger than $ r_{kki\tilde{i}}$, and (c) is due to the first-order Taylor approximation, $e^{a} \approx 1 + a$ and (\ref{pdf}). Moreover, (e) is due to $\int e^{Ar^2}rdr = \frac{e^{Ar^2}}{2A}$.
So, by substituting (\ref{Anew}) in (\ref{laplas}), we have:
\begin{align}
F_{\tilde{S}}(\tilde{s})=L^{-1}\{\frac{1}{s}\frac{\alpha-1}{s+\alpha-1}\}=1-e^{(1-\alpha)\tilde{s}}\label{laplas2}
\end{align}
Finally, by substituting (\ref{laplas2}) into (\ref{F_Snew}), $R_0$ and $R_1$ are derived as Lemma \ref{lemma2}.

\textbf{Derivation of the lower bounds:
} The CDF of \(\tilde{S}_0\) is \(F_{\tilde{S}_0}(s_0) = 1 - e^{2(1-\alpha)s_0}\) (meaning \(\tilde{S}_0\) follows an exponential distribution with rate \(\lambda = 2(\alpha - 1)\)). Since \(\log_2(1+\frac{a_0}{R b_0 \tilde{S}_0})\) is convex in \(\tilde{S}_0\) (its second derivative is positive), we apply Jensen's inequality, which states that for a convex function \(\phi\), we have \(\mathbb{E}[\phi(\tilde{S})] \geq \phi(\mathbb{E}[\tilde{S}])\). This yields a lower bound on the ER of the central user:  
\begin{align}
\mathbb{E}[\log_2(1+\frac{a_0}{R b_0 \tilde{S}_0})] \geq \log_2(1+\frac{a_0}{R b_0 \mathbb{E}[\tilde{S}_0]}). \label{Jens}
\end{align}
For \(\tilde{S}_0 \sim \text{Exp}(\lambda)\), \(\mathbb{E}[\tilde{S}_0] = \frac{1}{\lambda} = \frac{1}{2(\alpha-1)}\). Thus:  
$R_0 \geq \log_2\left(1 + \frac{2(\alpha-1)a_0}{R b_0}\right).$
\\
Moreover, from (\ref{F_Snew}) we have: $f_{\tilde{S}_1}(s_1) = \frac{d}{ds_1}F_{\tilde{S}_1}(s_1) = 2(\alpha-1)e^{(1-\alpha)s_1}\left(1 - e^{(1-\alpha)s_1}\right)$ (Note: $1-\alpha$ is negative since $\alpha > 1$). Thus, $\mathbb{E}[\tilde{S}_1] = \int_0^\infty s_1 \cdot f_{\tilde{S}_1}(s_1) \, ds_1 = 2(\alpha-1) \int_0^\infty s_1 e^{(1-\alpha)s_1} \left(1 - e^{(1-\alpha)s_1}\right) ds_1$. By letting $u = (\alpha-1)s_1$, we get:
\begin{align}
\mathbb{E}[\tilde{S}_1]&= \frac{2}{\alpha-1} \int_0^\infty u e^{-u} (1 - e^{-u}) du\nonumber\\
&= \frac{2}{\alpha-1} ( \int_0^\infty u e^{-u} du - \int_0^\infty u e^{-2u} du )\nonumber\\
&=\frac{2}{\alpha-1} ( \Gamma(2) - \frac{\Gamma(2)}{4} )= \frac{2}{\alpha-1} \cdot \frac{3}{4} = \frac{3}{2(\alpha-1)}
\end{align}
Moreover, $\log_2(1 + \frac{a_1}{\frac{a_0 b_1}{b_0} + R b_1 \tilde{S}_1})$ is convex where the proof is as follows. Let $c = \frac{a_0 b_1}{b_0}$, $d = R b_1$, and define $\phi(\tilde{S}_1) = \log_2(1 + \frac{a_1}{c + d\tilde{S}_1})$. The second derivative is $\phi''(\tilde{S}_1) = \frac{a_1 d^2 (2c + 2d\tilde{S}_1 + a_1)}{(c + d\tilde{S}_1)^2 (c + d\tilde{S}_1 + a_1)^2 \ln 2} > 0$. Since all terms are positive, $\phi(\tilde{S}_1)$ is convex in $\tilde{S}_1$. Thus, by applying Jensen's Inequality and substituting $\mathbb{E}[\tilde{S}_1] = \frac{3}{2(\alpha-1)}$, a lower bound on the ER of the second user is $R_1 \geq \log_2(1 + \frac{a_1}{\frac{a_0 b_1}{b_0} + \frac{3 R b_1}{2(\alpha - 1)}})$.
\section{Proof of Lemma \ref{lemma3}}\label{lemma3p}
According to statistical properties, we have:
\begin{align}
E[\log(1+\mathrm{SINR}^{w_0}_{\tilde{e}})]=\frac{1}{\ln{2}}\int_{0}^{\infty}\frac{1-F_{\mathrm{SINR}^{w_0}_{\tilde{e}}}(x)}{1+x}dx.\label{R_e}
\end{align}
Here, $F_{\mathrm{SINR}^{w_0}_{\tilde{e}}}(x)$ represents the CDF of the eavesdropper's SINR when eavesdropping signals from the central user. By considering the attacker with the best channel condition (i.e., worst case scenario), using (\ref{jam}) and defining $I= \sum_{l\neq k}^{\infty}r^{-2\alpha}_{l\tilde{e}} $, we obtain:
\begin{align}
&F_{\mathrm{SINR}^{w_0}_{\tilde{e}}}(x)=P(\max_{\tilde{e} \in \mathrm{\phi}_\text{e}}\frac{\frac{d_{\tilde{i}}a_0r^{-2\alpha}_{k\tilde{e}}}{b_
0}}{r^{-2\alpha}_{k\tilde{e}}(R-\frac{d_{\tilde{i}}a_0}{b_0})+R\sum_{l\neq k}^{\infty}r^{-2\alpha}_{l\tilde{e}}}<x)\nonumber\\
&\overset{(a)}{=}E_{\mathrm{\phi}_\text{e}}[\prod_{\tilde{e} \in \mathrm{\phi}_\text{e}}^{}P(\frac{\frac{d_{\tilde{i}}a_0r^{-2\alpha}_{k\tilde{e}}}{b_
0}}{r^{-2\alpha}_{k\tilde{e}}(R-\frac{d_{\tilde{i}}a_0}{b_0})+RI}<x|\mathrm{\phi}_\text{e})]\nonumber\\
&\overset{(b)}{=}\exp(-2 \pi \lambda_e \int_{0}^{\infty}(1-P(\frac{\frac{d_{\tilde{i}}a_0r^{-2\alpha}}{b_
0}}{r^{-2\alpha}(R-\frac{d_{\tilde{i}}a_0}{b_0})+RI}<x))rdr.\nonumber\\&
\overset{(c)}=\exp(-2 \pi \lambda_e[\sum_{u=1}^{U}(-1)^{u+1}{U\choose{u}}\frac{e^{-xu\tilde{\eta} (\frac{Rb_0}{d_{\tilde{i}}a_0}-1)}}{2(\frac{\pi \lambda_b x u \tilde{\eta} R b_0}{(\alpha-1)d_{\tilde{i}}a_0})}]),\label{fsinre}
\end{align}
where $(a)$ is due to maximum statistical property, $(b)$ is due to the PGFL of the PPP, and (c) is due to following the same approach as (\ref{ea}), utilize the approximation $e^{a} \approx 1 + a$, $\int e^{Ar^2}rdr = \frac{e^{Ar^2}}{2A}$, defining $\hat{\eta}\triangleq U(U!)^\frac{-1}{U} $, and the PGFL of the PPP and using the fact that by conditioning on the distance between the attacker and the serving BS in the $k$-th cell, the point process of other BSs, i.e., $\mathrm{\phi}_\text{b}\setminus k$, follows a reduced Palm distribution of the PPP where,  by the use of the Slivnyak-Mecke theorem,  is the same as the original PPP \cite{book}. In brief, first, we condition the event on the attacker's location. Then, we average over BS locations. 

Finally, by substituting (\ref{fsinre}) into (\ref{R_e}), the ergodic leakage rate to the attacker when eavesdropping the central user signal is shown by (\ref{e55}) to complete the proof.

\textbf{An approximation of the leakage to the eavesdropper:} By using the exponential approximation \cite{atractableapproach} for interference in PPP: $P(I > y) \approx e^{-\pi \lambda_b y^{-1/\alpha} \Gamma(1 + 1/\alpha)}$, and by substituting \( y = \frac{r^{-2\alpha} (C - x (R - C))}{x R} \), where $C = \frac{d_{\tilde{i}} a_0}{b_0}$, the inner probability at (\ref{fsinre}) (b) is approximated as: $P(I > \frac{r^{-2\alpha} (C - x (R - C))}{x R}) \approx \exp(-\pi \lambda_b (\frac{x R}{C - x (R - C)})^{1/\alpha} r^2 \Gamma(1 + 1/\alpha)).$ By letting \( \beta = \pi \lambda_b (\frac{x R}{C - x (R - C)})^{1/\alpha} \Gamma(1 + 1/\alpha) \), the integral at (\ref{fsinre}) (b) simplifies to: $\int_0^\infty (1 - e^{-\beta r^2}) r \, dr = \frac{1}{2 \beta}.$ Thus:
$F_{\mathrm{SINR}^{w_0}_{\tilde{e}}}(x) \approx \exp(-\frac{\pi \lambda_e}{\beta}) = \exp(-\frac{\lambda_e}{\lambda_b}(\frac{C - x (R - C)}{x R})^{1/\alpha} \frac{1}{\Gamma(1 + 1/\alpha)}).$ Thus, by letting \( \bar{\eta} = \frac{d_{\tilde{i}} a_0}{R b_0} \) and substituting \( F_{\mathrm{SINR}^{w_0}_{\tilde{e}}}(x) \), an approximation of (\ref{R_e}) is derived.
\section{Proof of Lemma \ref{lemma1new}}\label{lemma1newp}
As discussed in (\ref{model}), we assume that the desired user is located at y and the serving BS is at the origin. Also, we let $R_m=\min_{x \in \mathrm{\Phi}_\text{b}}\parallel x-y\parallel$ and $x_m=argmin_{x \in \mathrm{\Phi}_\text{b}}\parallel x-y\parallel$. Now, we obtain the CDF of $S_0$ and $S_1$:
\begin{align}
\!\!1\!-\!F_{S_0}(s_0)\!\!&\overset{(a)}=\!\!\sum_{n=1}^{N}\!\!(-1)^{n+1}\!{N\choose{n}}\!E[e^{-s_0n\eta r_{kk0\tilde{i}}^{2\alpha}\sum_{l\neq  k}^{}r_{lk0\tilde{i}}^{-2\alpha}
}],\label{F0new}
\end{align}
(a) follows from the similar approach as in (\ref{ea}). Moreover, we have:
\begin{align}
&E_{\phi,R_{kk0\tilde{i}}}[e^{-s_0n\eta r_{kk0\tilde{i}}^{2\alpha}\sum_{l\neq  k}^{}r_{lk0\tilde{i}}^{-2\alpha}
}]\nonumber\\
&\overset{(a)}{=}\!\!\!E_{\!R_m,R_{kk0\tilde{i}}}\!\![e^{-\eta ns_0r_{kk0\tilde{i}}^{2\alpha}r_m^{-2\alpha}\!\!+\!\!\int_{r_m}^{}\!\![e^{-\eta ns_0x^{-2\alpha}r_{kk0\tilde{i}}^{2\alpha}}-1]2\pi \lambda_b xdx}]\nonumber\\
&\overset{(b)}=\frac{(2\pi \rho \lambda_b)^2}{\tau^2}\!\!\!\!\int_{0}^{\infty}\!\!\!\!\!\int_{0}^{\tau r_m}\!\!\!\!\!\!\!\!\!\!W_0(r_m,r_{kk0\tilde{i}})dr_mdr_{kk0\tilde{i}},\label{E0new}
\end{align}
where (a) is due to approximating $\mathrm{\Phi}_\text{b} \setminus x_m$ with a PPP with the original density, $\lambda_b$, and using PGFL of the PPP \cite{book} and the fact that distances from other BSs to the user must be larger than $ r_m$. (b) is due to using the first order Taylor approximation \(e^a \approx 1 + a\) and the joint PDF of $R_m$ and $R_{kk0\tilde{i}}$, which can be derived using the similar approach as in \cite{new1}, given by $f_{R_{kk0\tilde{i}},R_m}(r_{kk0\tilde{i}},r_m)=\frac{(2\pi \rho \lambda_b)^2}{\tau^2} r_{kk0\tilde{i}}r_m\exp(-\pi \rho \lambda_b r_m^2)$. We rewrite the exponent $E$ in $W_0(r_m, r_{kk0\tilde{i}}) = r_m r_{kk0\tilde{i}} \exp(-\pi \rho \lambda_b r_m^2 - \eta n s_0 r_{kk0\tilde{i}}^{2\alpha} r_m^{-2\alpha} + \frac{\pi \lambda_b \eta n s_0 r_{kk0\tilde{i}}^{2\alpha} r_m^{-2\alpha + 2}}{1 - \alpha})$ by letting $A = \pi \rho \lambda_b$, $B = \eta n s_0$, $C = \frac{\pi \lambda_b}{1 - \alpha}$. Thus:
\[
E = -A r_m^2 + B r_{kk0\tilde{i}}^{2\alpha} r_m^{-2\alpha} (-1 + C r_m^2)
\]
By performing the substitution $u = r_m$, $v = \frac{r_{kk0\tilde{i}}}{u} \implies r_{kk0\tilde{i}} = u v$, $dr_{kk0\tilde{i}} = u \, dv$, the limits at (\ref{E0new}) become $v \in [0, \tau]$ and $u \in [0, \infty)$. The Jacobian is $u$, so $dr_m \, dr_{kki\tilde{i}} = u \, du \, dv$. Now express $W_0$ in $(u,v)$:
\[
W_0 = u^2 v \exp\left((-A + B C v^{2\alpha}) u^2 - B v^{2\alpha}\right).
\]
The double integral becomes:
\[
\int_{0}^{\infty} \int_{0}^{\tau} u^3 v \exp\left((-A + B C v^{2\alpha}) u^2 - B v^{2\alpha}\right) dv \, du.
\]
First solve the inner integral over $u$: $I = \int_{0}^{\infty} u^3 \exp\left(-D u^2\right) du$ where $D = A - B C v^{2\alpha}$ (assuming $D > 0$). Let $x = u^2$, $dx = 2u \, du$:
\begin{align}
I = \frac{1}{2} \int_{0}^{\infty} x e^{-D x} dx = \frac{1}{2D^2}.\label{I}
\end{align}
Thus, we have:
\begin{align}
\int_{0}^{\infty}\!\!\!\!\!\int_{0}^{\tau r_m}\!\!\!\!\!\!\!\!\!\!W_0(r_m,r_{kk0\tilde{i}})dr_mdr_{kk0\tilde{i}}&=\int_{0}^{\tau} \frac{v e^{-B v^{2\alpha}}}{2(A - B C v^{2\alpha})^2} dv\nonumber\\
&\overset{(a)}{=}\frac{1}{4\alpha} \int_{0}^{\tau^{2\alpha}} \frac{w^{\frac{1-\alpha}{\alpha}} e^{-B w}}{(A - B C w)^2} dw,\label{simpleintegral1}
\end{align}
where (a) is due to $w = v^{2\alpha}$, $dw = 2\alpha v^{2\alpha-1} dv$:
Thus according to (\ref{F0new}), (\ref{E0new}), and (\ref{simpleintegral1}) the CDF of $S_0$ is shown at (\ref{cdfs0new}).
 Moreover, by using the joint PDF of $R_m$ and $R_{kk1\tilde{i}}$ given as $f_{R_{kk1\tilde{i}},R_m}(r_{kk1\tilde{i}},r_m)=\frac{(2\pi \rho \lambda_b)^2}{1-\tau^2} r_{kk1\tilde{i}}r_m\exp(-\pi \rho \lambda_b r_m^2)$ (\cite{new1}) and using similar approach as the derivation of \ref{E0new}, we can derive $E_{\phi,R_{kk1\tilde{i}}}[e^{-s_1n\eta r_{kk1\tilde{i}}^{2\alpha}\sum_{l\neq  k}^{}r_{lk1\tilde{i}}^{-2\alpha}
}]$, and use it to derive $F_{S_1}(s_1)$ as similar approach to (\ref{F0new}), (\ref{E0new}), and (\ref{simpleintegral1}), resulting in (\ref{cdfs1new}). Due to space limitations, the
detailed steps are omitted.
\section{Proof of Lemma \ref{lemma2new}}\label{lemma2newp}
In the following, we provide two distinct approaches for deriving the ERs of the
central and second users. Then, we derive lower bounds on these ERs.

\textbf{Method 1:} Following the same approach as in (\ref{R_d}) and (\ref{R_d2}), The ER of the central and second user is $
R_{a0}=\int_{0}^{\infty}(1-F_{S_0}(\frac{TRb_0}{a_0}))dt$ and $
R_{a1}=\int_{0}^{\log{\frac{a_1b_0}{a_0b_1}+1}}(1-F_{S_1}(\frac{TRb_1}{a_1-\frac{Ta_0b_1}{b_0}}))dt$. By substituting (\ref{cdfs1new}) and (\ref{cdfs0new}) into them, the proof is complete.

\textbf{Method 2:} The CDF of \(\tilde{S_0}\), as defined in Method 2 in Appendix \ref{lemma2p}, is given by:
\begin{align}
&F_{\tilde{S_0}}(\tilde{s_0})=L^{-1}\{\frac{1}{s}E[e^{-s\tilde{S_0}}]\}
\overset{(a)}{=}\frac{(2\pi \rho \lambda_b)^2}{\tau^2}\!\!\nonumber\\
&\int_{0}^{\infty}\!\!\!\!\!\int_{0}^{\tau r_m}\!\!\!\!\!\!\!\!\!\!\!\!r_{kk0\tilde{i}}r_me^{-\pi \rho \lambda_b r_m^2}\!L^{-1}\!\{\frac{1}{s}e^{-sr_{kk0\tilde{i}}^{2\alpha}r_m^{-2\alpha}(1-\frac{\pi \lambda_b r_m^{2}}{1-\alpha})}\!\}dr_m\!dr_{kk0\tilde{i}},\nonumber\\
&\overset{(b)}{=}\!\!\frac{(2\pi \rho \lambda_b)^2}{\tau^2}\!\!\!\!\int_{0}^{\infty}\!\!\!\!\!\!\!r_me^{-\pi \rho \lambda_b r_m^2}(\!\!\int_{0}^{\tau r_m}\!\!\!\!\!\!\!\!\!\!U_{c}(\tilde{s_0})r_{kk0\tilde{i}}dr_{kk0\tilde{i}})dr_m,\nonumber\\
&\overset{(c)}=\!\!\frac{(2\pi \rho \lambda_b)^2}{2\tau^2}\!\!\!\int_{0}^{\infty}\!\!\!\!\!\!\!r^3_me^{-\pi \rho \lambda_b r_m^2}\!\min( \tilde{s_0}^{1/\alpha} (1 - \frac{\pi \lambda_b r_m^2}{1-\alpha})^{-1/\alpha}\!\!\!,\!\tau^2 )dr_m
\label{laplas2new}
\end{align}
where (a) is due to the linearity of the Laplace inverse and the substitution of \(\eta n s_0\) in (\ref{E0new}) with \(s\), (b) follows from defining $c\triangleq r_{kk0\tilde{i}}^{2\alpha}r_m^{-2\alpha}\!\!(1-\frac{\pi \lambda_b r_m^{2}}{1-\alpha})$ and \(L^{-1}\left\{\frac{1}{s}e^{-cs}\right\} = U_c(t)\), where \(U_c(t)\) is the shifted unit step function, (c) is because \(U_c(t)=1\) when $t>c$ and then evaluating the integral.
Thus, by substituting (\ref{laplas2new}) (b) in $R_{a0}=\!\!\int_{0}^{\infty}\!\!(F_{\tilde{S_0}}(\frac{a_0}{(2^t-1)Rb_0}))dt$, we have:
\begin{align}
R_{a0}\!
&\overset{(a)}{=}\!\frac{(2\pi \rho \lambda_b)^2}{\tau^2}\!\!\!\int_{0}^{\infty}\!\!\!\!\!\int_{0}^{\tau r_m}\!\!\!\!\!\!\!\!\!\!\!r_mr_{kk0\tilde{i}}e^{-\pi \rho \lambda_b r_m^2}\tilde{N}(r_{kk0\tilde{i}},r_m)dr_mdr_{kk0\tilde{i}}, \label{r0newlap}
\end{align}
where (a) is because of the change in the order of integration and using the shifted step function which is \(1\) when \(\frac{a_0}{(2^t-1)Rb_0}>r_{kk0\tilde{i}}^{2\alpha}r_m^{-2\alpha}(1-\frac{\pi \lambda_b r_m^{2}}{1-\alpha})\).

\textbf{Derivation of the lower bound of $R_{a0}$:} 
From (\ref{Jens}), we have $R_{a0} \geq \log_2\left(1 + \frac{a_0}{R b_0 \mathbb{E}[\tilde{S_0}]}\right)$. To compute this, we need to derive $\mathbb{E}[\tilde{S_0}]$ using the CDF $F_{\tilde{S_0}}(\tilde{s_0})$ given at (\ref{laplas2new}). The transition point of the integral at (\ref{laplas2new}) occurs when $\tilde{s_0}^{1/\alpha} \psi(r^*) = \tau^2$, where $\psi(r_m) = \left(1 - \frac{\pi \lambda_b r_m^2}{1-\alpha}\right)^{-1/\alpha}$, yielding: $r^* = \sqrt{\frac{1-\alpha}{\pi \lambda_b} \left(1 - \frac{\tilde{s_0}}{\tau^{2\alpha}}\right)}.$ \( r^* \) is real only if \( \tilde{s_0} \geq \tau^{2\alpha} \). For \( \tilde{s_0} < \tau^{2\alpha} \), \( \min(\cdot) = \tilde{s_0}^{1/\alpha} \psi(r_m) \) for all \( r_m \). Thus, the CDF splits into two cases:
1) \( \tilde{s_0} \geq \tau^{2\alpha} \):
  $ F_{\tilde{S_0}}(\tilde{s_0}) = \frac{(2\pi \rho \lambda_b)^2}{2\tau^2} [ \tau^2 \int_0^{r^*} r_m^3 e^{-c r_m^2} dr_m + \tilde{s_0}^{1/\alpha} \int_{r^*}^\infty r_m^3 e^{-c r_m^2} \psi(r_m) dr_m].$
2) \( \tilde{s_0} < \tau^{2\alpha} \):
   $F_{\tilde{S_0}}(\tilde{s_0}) = \frac{(2\pi \rho \lambda_b)^2}{2\tau^2} \tilde{s_0}^{1/\alpha} \int_0^\infty r_m^3 e^{-c r_m^2} \psi(r_m) dr_m.$
Moreover, $\mathbb{E}[\tilde{S_0}] = \int_0^\infty (1 - F_{\tilde{S_0}}(s)) ds.$ For \( s \geq \tau^{2\alpha} \), e.i. for large \( s \), \( F_{\tilde{S_0}}(s) \approx 1 \), so \( 1 - F_{\tilde{S_0}}(s) \approx 0 \).  For \( s < \tau^{2\alpha} \), we have \( F_{\tilde{S_0}}(s) \approx \frac{(2\pi \rho \lambda_b)^2}{2\tau^2} s^{1/\alpha} \int_0^\infty r_m^3 e^{-c r_m^2} \psi(r_m) dr_m \).  
Let \( I = \int_0^\infty r_m^3 e^{-c r_m^2} \psi(r_m) dr_m \). Since \( \psi(r_m) \) decays as \( r_m \to \infty \), approximate \( \psi(r_m) \approx 1 \) for small \( r_m \):
\[
I \approx \int_0^\infty r_m^3 e^{-c r_m^2} dr_m = \frac{1}{2c^2}.
\]
Thus: $F_{\tilde{S_0}}(s) \approx \frac{(2\pi \rho \lambda_b)^2}{2\tau^2} s^{1/\alpha} \cdot \frac{1}{2c^2} = \frac{s^{1/\alpha}}{\tau^2}.$
Now, $\mathbb{E}[\tilde{S_0}] \approx \int_0^{\tau^{2\alpha}}(1 - \frac{s^{1/\alpha}}{\tau^2}) ds = \tau^{2\alpha} (1 - \frac{1}{1 + 1/\alpha}) = \frac{\tau^{2\alpha}}{1 + \alpha}.$ Thus, by substituting \( \mathbb{E}[\tilde{S_0}] \approx \frac{\tau^{2\alpha}}{1 + \alpha} \) into Jensen’s inequality:
$R_{a0} \geq \log_2(1 + \frac{a_0 (1 + \alpha)}{R b_0 \tau^{2\alpha}}).$

Moreover, by substituting \(\eta n s_1\) with \(s\) in \(E_{\phi,R_{kk1\tilde{i}}}[e^{-s_1 n \eta r_{kk1\tilde{i}}^{2\alpha} \sum_{l \neq k} r_{lk1\tilde{i}}^{-2\alpha}}]\) given in Appendix \ref{lemma1newp}, and following the same approach as in (\ref{laplas2new}), the CDF of \(\tilde{S_1}\) is derived as:
\begin{align}
&F_{\tilde{S_1}}(\tilde{s_1})\overset{(b)}{=}\frac{(2\pi \rho \lambda_b)^2}{1-\tau^2}\!\!\int_{0}^{\infty}\!\!\!\int_{\tau r_m}^{ r_m}r_mr_{kk1\tilde{i}}e^{-\pi \rho \lambda_b r_m^2}\nonumber\\&U_{r_{kk1\tilde{i}}^{2\alpha}r_m^{-2\alpha}(1-\frac{\pi \lambda_b r_m^{2}}{1-\alpha})}(\tilde{s_1})dr_mdr_{kk1\tilde{i}}\nonumber\\
&\overset{(c)}{=}\!\frac{(2\pi \rho \lambda_b)^2}{2(1-\tau^2)}\!\!\!\int_{0}^{\infty}\!\!\!\!\!\!r^3_me^{-\pi \rho \lambda_b r_m^2}[\min( \tilde{s_1}^{1/\alpha} (1 - \frac{\pi \lambda_b r_m^2}{1-\alpha})^{-1/\alpha},1 )\nonumber\\&-\min( \tilde{s_1}^{1/\alpha} (1 - \frac{\pi \lambda_b r_m^2}{1-\alpha})^{-1/\alpha},\tau^2 )]
dr_m.\label{laplas3}
\end{align}
Thus, by substituting (\ref{laplas3}) (b) in $R_{a1}=\int_{0}^{\log{\frac{a_1b_0}{a_0b_1}+1}}(F_{\tilde{S_1}}(\frac{a_1-\frac{Ta_0b_1}{b_0}}{TRb_1}))dt$, we have:
\begin{align}
R_{a1}\!\!
&\overset{(a)}{=\!}\!\frac{(2\pi \rho \lambda_b)^2}{1-\tau^2}\!\!\!\int_{0}^{\infty}\!\!\!\!\!\int_{\tau r_m}^{r_m}\!\!\!\!\!\!\!\!\!r_mr_{kk1\tilde{i}}e^{-\pi \rho \lambda_b r_m^2}\tilde{M}(r_{kk1\tilde{i}},r_m)dr_mdr_{kk1\tilde{i}}, \label{r1newlap}
\end{align}
(a) is because of the change in the order of integration and the nonzero range of the shifted step function.

\textbf{Derivation of the lower bound of $R_{a1}$:} 
Following similar methodology to the RP case, we have: $R_{a1} \geq \log_2(1 + \frac{a_1}{\frac{a_0 b_1}{b_0} + R b_1 \mathbb{E}[\tilde{S_1}]}).$ We derive $\mathbb{E}[\tilde{S_1}]$ using the CDF $F_{\tilde{S_1}}(\tilde{s_1})$ derived at (\ref{laplas3}). The CDF integrand splits into three regimes:
1. Far-field ($\tilde{s_1}^{1/\alpha} \psi(r_m) \geq 1$): Contribution $(1 - \tau^2)$
2. Mid-field ($\tau^2 \leq \tilde{s_1}^{1/\alpha} \psi(r_m) < 1$): Contribution $(\tilde{s_1}^{1/\alpha} \psi(r_m) - \tau^2)$
3. Near-field ($\tilde{s_1}^{1/\alpha} \psi(r_m) < \tau^2$): Contribution $0$.
\\
Transition points $r_1^*$ and $r_2^*$ satisfy:
$\tilde{s_1}^{1/\alpha} \psi(r_1^*) = \tau^2, \quad \tilde{s_1}^{1/\alpha} \psi(r_2^*) = 1,$ with solutions: $r^* = \sqrt{\frac{1-\alpha}{\pi \lambda_b} (1 - \tilde{s_1}^{-1/\alpha} \cdot \text{threshold}^{-\alpha})}, \quad \text{threshold} \in \{\tau^2, 1\}.$
\\
For $\tilde{s_1} \geq \tau^{2\alpha}$, the CDF approximates to: $F_{\tilde{S_1}}(\tilde{s_1}) \approx \frac{(2\pi \rho \lambda_b)^2}{2(1-\tau^2)}[ \int_{r_1^*}^{r_2^*} (\tilde{s_1}^{1/\alpha} \psi(r_m) - \tau^2) e^{-\pi \rho \lambda_b r_m^2} r_m^3 dr_m + \int_{r_2^*}^{\infty} (1 - \tau^2) e^{-\pi \rho \lambda_b r_m^2} r_m^3 dr_m ].$ Using approximations:
1. Near-field ($r_m \leq r_2^*$): $\psi(r_m) \approx 1$
2. Far-field ($r_m > r_2^*$): Dominated by $(1 - \tau^2)$. The expectation becomes: $\mathbb{E}[\tilde{S_1}] \approx \int_{0}^{\tau^{2\alpha}} 1 \, ds + \int_{\tau^{2\alpha}}^{1} (1 - \frac{s^{1/\alpha} - \tau^2}{1-\tau^2}) ds = \tau^{2\alpha} + \frac{\alpha}{\alpha + 1} (1 - \tau^{2(\alpha + 1)/\alpha}).$ Substituting into Jensen's inequality prove the lower bound given ta Lemma \ref{lemma2new}.
\section{Proof of Lemma \ref{lemma3new}}\label{lemma3newp}
In the following, we present two distinct approaches for deriving the ergodic rates (ERs) of the central and second users, along with their corresponding lower bounds.

\textbf{Method 1:} If $S_0<S_1$, then $ \mathrm{SINR}^{w_1}_{k0\tilde{i}} < \mathrm{SINR}^{w_1}_{k1\tilde{i}} $ and the central user can not perform SIC. Thus, the SINR of the central user in decoding its own message is the same as calculating (\ref{hello}) but we must not omit the term caused by the second user signal which acts as an interference and shows itself in the denominator. After some simple operation we obtain: $\mathrm{SINR}^{w_0}_{k0\tilde{i}}=\frac{a_0r^{-2\alpha}_{kk0\tilde{i}}}{Rb_0\sum_{l\neq k}^{\infty}r^{-2\alpha}_{lk0\tilde{i}}+\frac{a_1b_0r^{-2\alpha}_{kk0\tilde{i}}}{b_1}}$. In addition, we have: $ \mathrm{SINR}^{w_0}_{k1\tilde{i}} > \mathrm{SINR}^{w_0}_{k0\tilde{i}} $, which $ \mathrm{SINR}^{w_0}_{k1\tilde{i}}=\frac{\frac{b_1a_0r^{-2\alpha}_{kk1\tilde{i}}}{b_0}}{Rb_1\sum_{1\neq k}^{\infty}r^{-2\alpha}_{lk1\tilde{i}}+a_1r^{-2\alpha}_{kk1\tilde{i}}}$ is the SINR of the second user in decoding the central user message. This means that the second user can perform SIC and omit the central user signal. Therefore, the SINR of the second user after SIC is the same as (\ref{hello2}) except for one term in the denominator which indicates the interference caused by the central user. Thus, we have $\mathrm{SINR}^{w_1}_{k1\tilde{i}}=\frac{a_1r^{-2\alpha}_{kk1\tilde{i}}}{Rb_1\sum_{l\neq k}^{\infty}r^{-2\alpha}_{lk1\tilde{i}}}$. Then, following the same approach as (\ref{R_d2}) and (\ref{R_d}), we have $\!\!R_{b0}\!\!=\!\!\!\int_{0}^{\log{\frac{a_0b_1}{a_1b_0}+1}}\!\!\!\!\!\!\!\!\!\!\!\!\!\!\!\!\!\!\!\!\!(1-F_{S_0}(\!\frac{TRb_0}{a_0-\frac{Ta_1b_0}{b_1}}))dt$ and $R_{b1}=\int_{0}^{\infty}(1-F_{S_1}(\frac{TRb_1}{a_1}))dt$. Thus, by substituting (\ref{cdfs1new}) and (\ref{cdfs0new}), the proof is complete.

\textbf{Method 2:} By the definition of $\tilde{S_0}$ and $\tilde{S_1}$ as given in Appendix (\ref{lemma2p}), we have $R_{b0}=\!\!\int_{0}^{\log{\frac{a_0b_1}{a_1b_0}+1}}\!\!(F_{\tilde{S_0}}(\frac{a_0-\frac{(2^t-1)a_1b_0}{b_1}}{(2^t-1)Rb_0}))dt$ and $R_{b1}=\int_{0}^{\infty}(F_{\tilde{S_1}}(\frac{a_1}{(2^t-1)Rb_1}))dt$. Then, by replacing $F_{\tilde{S_0}}(\tilde{s_0})$ and $F_{\tilde{S_1}}(\tilde{s_1})$ derived at (\ref{laplas2new}) (b) and (\ref{laplas3}) (b) and following the same approach as (\ref{r0newlap}) and (\ref{r1newlap}), the proof is complete. Due to space limitations, we do not provide the details.

\textbf{Derivation of the lower bound of $R_{b0}$ and $R_{b1}$:} 
Using the previously derived expressions for $E[\tilde{S_0}]$ and $E[\tilde{S_1}]$, provided in Appendix \ref{lemma2newp}, along with the new SINR expressions introduced in Method 1 of this appendix, and applying Jensen's inequality, the proof is complete. Due to space limitations, the detailed steps are omitted.
\section{Proof of Lemma \ref{lemma4}}\label{lemma4p}
We propose two method for deriving $P^{w_0}_{\mathrm{out}}$ and $P^{w_1}_{\mathrm{out}}$.

\textbf{Method 1:} $P^{w_0}_{\mathrm{out}}$ is expressed as follows:
\begin{align}
&P^{w_0}_{\mathrm{out}}\!\!=\!P(\log(1+\mathrm{SINR}^{w_0}_{k0\tilde{i}})\!\!-\!\!\log(1+\mathrm{SINR}^{w_0}_{\tilde{e}})\!<\!\tilde{R}_0)\nonumber\\&\!\!\overset{(a)}{=}\!\!\!\!\int_{0}^{\infty}\!\!\!\!F_{S_0}(\frac{Rb_0(2^{\tilde{R}_0}(1+z)-1)}{a_0})f_{\mathrm{SINR}^{w_0}_{\tilde{e}}}(z)dz,
\label{pout11}
\end{align}
(a) is obtained by conditioning on the attacker SINR where $f_{\mathrm{SINR}^{w_0}_{\tilde{e}}}$ is the PDF of eaves SINR when eavesdropping the central user. Moreover, 
$F_{S_0}(\frac{Rb_0(2^{\tilde{R}_0}(1+z)-1)}{a_0})\overset{(a)}=(\sum_{n=0}^{N}(-1)^{n}{N\choose{n}}\frac{-\alpha+1}{-\alpha+1-\eta n(\frac{Rb_0(2^{\tilde{R}_0}(1+z)-1)}{a_0})})^2$
where (a) is obtained by substituting (\ref{F_Sfinal}) into (\ref{F_S}) and replacing $s$ by $\frac{Rb_0(2^{\tilde{R}_0}(1+z)-1)}{a_0}$. $f_{\mathrm{SINR}^{w_0}_{\tilde{e}}}$ is expressed by taking the derivative of $F_{\mathrm{SINR}^{w_0}_{\tilde{e}}}(x)$, given at (\ref{fsinre}), with respect to $x$. Thus, $f_{\mathrm{SINR}^{w_0}_{\tilde{e}}}(x)=F_{\mathrm{SINR}^{w_0}_{\tilde{e}}}(x) (\sum_{u=1}^{U}\!\!\!\!\!\!\!\frac{(-1)^{u+1}{U\choose{u}}(2 \pi \lambda_e e^{-xu\tilde{\eta} (\frac{Rb_0}{d_{\tilde{i}}a_0}-1)}\!\!\!\!\!)(xu\tilde{\eta} (\frac{Rb_0}{d_{\tilde{i}}a_0}-1)+1)}{2(\frac{\pi \lambda_b x^2 u \tilde{\eta} R b_0}{(\alpha-1)d_{\tilde{i}}a_0})})$.
Finally, $P^{w_0}_{\mathrm{out}}$ is expressed as shown in (\ref{pout13}).
Moreover, $P^{w_1}_{\mathrm{out}}\!\!=\!\!\!\!\int_{0}^{\infty}\!\!\!F_{S_1}\big(\frac{(2^{\tilde{R}_1}(1+z)-1)Rb_1}{a_1-\frac{(2^{\tilde{R}_1}(1+z)-1)a_0b_1}{b_0}}\big)f_{\mathrm{SINR}^{w_1}_{\tilde{e}}}(z)dz$ where $f_{\mathrm{SINR}^{w_1}_{\tilde{e}}}$ is the PDF of eaves SINR when eavesdropping the second user and is obtained by replacing $a_0$ and $b_0$ at $f_{\mathrm{SINR}^{w_0}_{\tilde{e}}}$ by $a_1$ and $b_1$.
If \( a > \max(\mathrm{SINR}^{w_1}_{k1\tilde{i}}) = \frac{a_1b_0}{a_0b_1} \), we have \( P(\mathrm{SINR}^{w_1}_{k1\tilde{i}} < a) = 1 \). Therefore, we need to consider two situations: If $0>(\frac{a_1b_0}{a_0b_1}+1)2^{-\tilde{R_1}}-1$, $P^{w_1}_{\mathrm{out}}=1$ otherwise:
\begin{align}
&P^{w_1}_{\mathrm{out}}\!\!=\!\!\!\!\int_{0}^{(\frac{a_1b_0}{a_0b_1}+1)2^{-\tilde{R_1}}-1}\!\!\!\!\!\!\!\!\!\!\!\!\!\!\!\!F_{S_1}\big(\frac{(2^{\tilde{R}_1}(1+z)-1)Rb_1}{a_1-\frac{(2^{\tilde{R}_1}(1+z)-1)a_0b_1}{b_0}}\big)f_{\mathrm{SINR}^{w_1}_{\tilde{e}}}(z)dz\nonumber\\&+\int_{(\frac{a_1b_0}{a_0b_1}+1)2^{-\tilde{R_1}}-1}^{\infty}\!\!\!\!f_{\mathrm{SINR}^{w_1}_{\tilde{e}}}(z)dz,\label{pout21con1}
\end{align}
By substituting (\ref{F_Sfinal}) into (\ref{F_S}) and replacing $s$ by $\frac{(2^{\tilde{R}_1}(1+z)-1)Rb_1}{a_1-\frac{(2^{\tilde{R}_1}(1+z)-1)a_0b_1}{b_0}}$, $P^{w_1}_{\mathrm{out}}$ is derived as (\ref{pout13}).

\textbf{Method 2:}
Given the definitions of \(\tilde{S_0}\) and \(\tilde{S_1}\) in Appendix \ref{lemma2p}, we have: $P^{w_0}_{\mathrm{out}}\!\!=\!\!\!\!\!\!\int_{0}^{\infty}\!\!\!\!(1-F_{\tilde{S_0}}(\frac{a_0}{Rb_0(2^{\tilde{R}_0}(1+z)-1)}))f_{\mathrm{SINR}^{w_0}_{\tilde{e}}}(z)dz$. Moreover, $F_{\tilde{S_0}}(\frac{a_0}{Rb_0(2^{\tilde{R}_0}(1+z)-1)})\overset{(a)}=1-e^{2(1-\alpha)\frac{a_0}{Rb_0(2^{\tilde{R}_0}(1+z)-1)}}$ where (a) is due to (\ref{F_Snew}) and (\ref{laplas2}) and by substituting $\tilde{s}$ by $\frac{a_0}{Rb_0(2^{\tilde{R}_0}(1+z)-1)}$. Thus, by using $f_{\mathrm{SINR}^{w_0}_{\tilde{e}}}$, $P^{w_0}_{\mathrm{out}}$ is derived as (\ref{pout13}).
In addition, we have: $P^{w_1}_{\mathrm{out}}\!\!=\!\!\!\int_{0}^{\infty}\!\!\!\!(1-F_{\tilde{S_1}}(\frac{a_1-\frac{(2^{\tilde{R}_1}(1+z)-1)a_0b_1}{b_0}}{(2^{\tilde{R}_1}(1+z)-1)Rb_1})f_{\mathrm{SINR}^{w_1}_{\tilde{e}}}(z)dz$. Using (\ref{F_Snew}) and (\ref{laplas2}) and by substituting $\tilde{s}$ by $\frac{a_1-\frac{(2^{\tilde{R}_1}(1+z)-1)a_0b_1}{b_0}}{(2^{\tilde{R}_1}(1+z)-1)Rb_1}$. $P^{w_1}_{\mathrm{out}}$ is derived as (\ref{pout13}).
\end{document}